\documentclass[a4paper]{article}

\usepackage{amsmath,amssymb,amscd,graphicx}
\usepackage{amsthm,color}
\usepackage{array}

\numberwithin{equation}{section}
\newtheorem{theorem}{\bf Theorem}[section]

\newtheorem{proposition}[theorem]{\bf Proposition}

\newtheorem{corollary}[theorem]{Corollary}
\theoremstyle{remark}
\newtheorem{remark}{\bf Remark}[section]

\begin{document}

\title{Analytic Extension of the Birkhoff Normal Forms for the Free Rigid Body Dynamics on $SO(3)$}
\author{Jean-Pierre Fran\c{c}oise\thanks{Laboratoire Jacque-Louis Lions, 
Universit\'{e} Pierre-Marie Curie, 4 Pl. Jussieu, 75252 
Paris, France.  
E-mail: \texttt{jean-pierre.francoise@upmc.fr}}
\, and\, Daisuke Tarama\thanks{Department of Mathematics, Kyoto University, Kitashirakawa-Oiwake-cho, Sakyo-ku, 606-8502, Kyoto, Japan. \quad JSPS Research Fellow. 
E-mail: \texttt{tarama@math.kyoto-u.ac.jp}}}
\date{This version: 20/July/2013}

\maketitle
\noindent\textbf{Key words} Birkhoff normal form, analytic extension, free rigid body, elliptic fibration, monodromy\\
\textbf{MSC(2010)}: \; 14D06, 37J35, 58K10, 58K50, 70E15 
\begin{center}Abstract\end{center}
Birkhoff normal form is a power series expansion associated with the local behavior of the Hamiltonian systems near a critical point. It is known to be convergent for integrable systems under some non-degeneracy conditions. By means of an expression of the inverse of Birkhoff normal form by a period integral, analytic continuation of the Birkhoff normal forms is considered for the free rigid body dynamics on $SO(3)$. The global behavior of the Birkhoff normal forms is clarified in relation to the monodromy of an elliptic fibration which naturally arises from the free rigid body dynamics. 

\section{Introduction}
In analytical mechanics, the motions of rigid bodies are basic problems. Among them, the free rigid body, which stands for the rigid body under no external force, is the simplest example. Its complete integrability and the stability of its equilibria are understood well through the long history of researches. In particular, geometric mechanics provides a well-organized description of this dynamical system. (See \cite{whittaker, abraham-marsden,marsden-ratiu}.) 

The free rigid body dynamics should first be defined as a Hamiltonian system on the cotangent bundle of the rotation group $SO(3)$. Because of the left-invariance of this system, it is essentially described by the so-called Euler equation posed on the angular momentum. Moreover, by means of the Marsden-Weinstein reduction, one can reduce the system onto the level surface of the norm of the angular momentum, which is a two-dimensional sphere. The reduced system is of one degree of freedom, and therefore completely integrable in the sense of Liouville. It is well known that there are generically six equilibria for the reduced systems on the sphere, four of which are stable and the other two are unstable. (cf. \cite{marsden-ratiu}.) 

Around a stationary point of a Hamiltonian system, it is possible to consider the normal form of the Hamiltonian. Historically, Birkhoff introduced the notion of Birkhoff normal forms as formal series and discussed the relation with the stability \cite{birkhoff_1,birkhoff_2}. It is known that the Birkhoff normal form of the Hamiltonian is convergent in a neighbourhood of the stationary point by a result of Vey \cite{vey} for analytic systems. The differentiable case of class $\mathcal{C}^{\infty}$ was studied by Eliasson \cite{eliasson}. The convergence of Birkhoff normal forms for analytic integrable systems in the degenerate case was shown by Ito \cite{ito} under the non-resonance condition. The convergence of the Birkhoff normal form for the systems with one degree of freedom was essentially shown by Siegel \cite{siegel-moser}.

The detailed structure of the  Birkhoff normal form has been studied recently for the pendulum \cite{francoise-garrido-gallavotti_1} and for the free rigid body \cite{francoise-garrido-gallavotti_2}. In \cite{francoise-garrido-gallavotti_2},  the Birkhoff normal forms both for the stable and unstable stationary points are considered by using the method of relative cohomology, and the properties of the Birkhoff normal forms and those of the inverse of the Birkhoff normal forms are discussed. 

On the other hand, the free rigid body dynamics is closely related to complex algebraic geometry because of its complete integrability. In fact, the integral curve of the free rigid body can be described as an intersection of two quadric surfaces in three-dimensional Euclidean space, which is a (real) elliptic curve. From the view point of complex algebraic geometry, it is natural to complexify and to compactify all the settings, which is also helpful to understand the deep geometric structure of the free rigid body dynamics. In view of this, some elliptic fibrations arising from the free rigid bodies have been considered in \cite{naruki-tarama}. In this paper, the fibrations are considered over the base space which includes not only the values of the Hamiltonian but also the principal axes of the inertia tensor, as their base coordinates. This even allows a geometric description of the bifurcation phenomena of the free rigid body dynamics as described in \cite{naruki-tarama}. The classification of the singular fibres of the elliptic fibrations is given in \cite{naruki-tarama} in relation to such a bifurcation phenomena. 

Although the Birkhoff normal form is by definition a local object associated to a stationary point for a Hamiltonian system, it is possible to enquire its analytic extension in the integrable case. In the present paper, the Birkhoff normal forms of the equilibria for the free rigid body dynamics and their analytic continuation are considered in relation to the naive elliptic fibration which is discussed in \cite{naruki-tarama}. The key to this relation is an expression of the derivative of the inverse for the Birkhoff normal form as a kind of period integral. This period integral has a close relationship to a special Gau{\ss} hypergeometric equation, which makes the concrete calculation rather easy. In particular, the global monodromy of this elliptic fibration is found by using the monodromy of the Gau{\ss} hypergeometric equation. 

The structure of the present paper is as follows: \\
In Section 2, some elements of the $SO(3)$ free rigid body are described. The Euler equation posed on the angular momentum and the reduction onto the sphere given as a level surfaces of the norm of the angular momentum are briefly described. For the reduced system, the equilibria and their stability are also explained. 

In Section 3, the relation of the Birkhoff normal form to some period integrals is discussed for an arbitrary Hamiltonian system of one degree of freedom. Indeed, the derivative of the inverse of the Birkhoff normal form can be expressed in terms of period integrals both around the elliptic and hyperbolic stationary points. 

Section 4 describes the application of the result in Section 3 to the case of the free rigid body dynamics. In this case, the explicit expression of the derivative of the inverse Birkhoff normal form is given in terms of the complete elliptic integral of first kind. From this expression, the period integral is connected to a special Gau{\ss}  hypergeometric equation. 

Section 5 deals with the relation between the result in Section 4 and the naive elliptic fibration arising from the free rigid bodies. The singular fibres of this elliptic fibration are classified in \cite{naruki-tarama}, but the global monodromy is not discussed there. In relation to this elliptic fibration, one finds that the above period integrals give rise to a basis of the first cohomology group of regular fibres. It is also to be noted that the period integral has some symmetry property, which makes the calculation easier. The results in this section are compared with the results of \cite{francoise-garrido-gallavotti_2}. In \cite[\S VI]{francoise-garrido-gallavotti_2}, it is already emphasised that the coefficients of the inverse for the Birkhoff normal form around an elliptic equilibrium on the axis corresponding to $I_3$ are symmetric polynomials $P_n$ in the parameter ${\displaystyle r^2=\frac{\frac{1}{I_1}-\frac{1}{I_2}}{\frac{1}{I_3}-\frac{1}{I_2}}}$, where $I_1$, $I_2$, $I_3$ stand for the principal axes of the inertia tensor of the rigid body, in the sense that 
\[
P_n(r^2)=\sum_{j=0}^nF_jr^{2j},\qquad F_j=F_{n-j}. 
\]
Note that it is assumed that $I_3<I_1<I_2$ in \cite{francoise-garrido-gallavotti_2}. It is shown here that this property is a consequence of a covariance of the analytic extensions of the derivative of the inverse Birkhoff normal forms relatively to the symmetry group ($\mathfrak{S}_4$) of the base space of the naive elliptic fibration arising from the free rigid body dynamics. This covariance is explained in the formulae \eqref{covariance}. On the top of this covariance formulae, another formula is derived in the spirit of the connection formula of the Gau{\ss} hypergeometric equation (cf. Proposition 5.4 and \cite{caratheodory}). 

Moreover, in Section 6, the calculation of the global monodromy of the elliptic fibration is performed, using monodromy and the connection formula of the Gau{\ss} hypergeometric equation. 

A similar study on the pendulum system might be performed and the relation to the rigid body dynamics might be clarified in view of \cite{holm-marsden} and \cite{iwai-tarama}.

\section{Free rigid body dynamics}
We start with a basic description of free rigid body dynamics. We denote the three-dimensional rotation group by $SO(3)$ and its Lie algebra by $\mathfrak{so}(3)$. Obviously, $\mathfrak{so}(3)$ consists of all the $3\times 3$ skew-symmetric matrices. The cotangent bundle $T^{\ast}SO(3)$ can be identified with $SO(3)\times {\mathfrak{so}(3)}^{\ast}$ by the left-translation: $T^{\ast}SO(3)\ni(g,\alpha_g)\mapsto (g,{L_g}^{\ast}\alpha_g)\in SO(3)\times {\mathfrak{so}(3)}^{\ast}$, where $\alpha_g\in T_g^{\ast}SO(3)$ and $L_g$ is the left-translation by $g\in SO(3)$ such as $L_g:SO(3)\ni a\mapsto ga\in SO(3)$. Using the natural projection $\pi: T^{\ast}SO(3)\rightarrow SO(3)$, we define the canonical one-form $\theta$ on $T^{\ast} SO(3)$ through $\theta\left. \left(\widetilde{X}\right)\right|_{(g,\alpha_g)}=\alpha_g\left(\pi_{\ast}\widetilde{X}\right)$ for $\widetilde{X}\in T_{(g,\alpha_g)} \left(T^{\ast}SO(3)\right)$. The canonical symplectic form $\Theta$ on $T^{\ast}SO(3)$ is defined by $\Theta=-\mathsf{d}\theta$. For a smooth function $\widetilde{H}$ on $T^{\ast}SO(3)$, the Hamiltonian vector field $X_{\widetilde{H}}$ with respect to the Hamiltonian $\widetilde{H}$ is given by $\mathsf{d}\widetilde{H}=\iota_{X_{\widetilde{H}}}\Theta$. Henceforth, we concentrate ourselves to the case where $\widetilde{H}$ is left-invariant, namely there is a function $H$ on ${\mathfrak{so}(3)}^{\ast}$ such that $\widetilde{H}\left(\left(g,\alpha_g\right)\right)=H\left({L_g}^{\ast}\alpha_g\right)$. Then, we have the following expression of $X_{\widetilde{H}}=\left(X_{\widetilde{H}}^{\prime},X_{\widetilde{H}}^{\prime\prime}\right)\in T_gSO(3)\times{\mathfrak{so}(3)}^{\ast}$. 
\begin{proposition}
The Hamiltonian vector field $X_{\widetilde{H}}$ is given by 
\begin{align}
X_{\widetilde{H}}^{\prime} & = {L_g}_{\ast}\left(\mathsf{d}H\right)_M, \notag \\
X_{\widetilde{H}}^{\prime\prime} & = \mathrm{ad}_{\left(\mathsf{d}H\right)_M}^{\ast}M, \notag
\end{align}
where $M=L_g^{\ast}\alpha_g\in{\mathfrak{so}(3)}^{\ast}$. 
\end{proposition}
For the proof, see \cite[Prop. 4.4.1., p. 315]{abraham-marsden}. 

To introduce the free rigid body, we consider the linear mapping $\mathcal{J}:\mathfrak{so}(3)\ni X\mapsto X\mathsf{J}+\mathsf{J}X\in\mathfrak{so}(3)$ defined in accordance with a $3\times 3$ symmetric matrix $\mathsf{J}$. On the Lie algebra $\mathfrak{so}(3)$, there is an invariant inner product $\langle\cdot,\cdot\rangle$ defined through $\langle X, Y\rangle=-\mathrm{Tr}\left(XY\right)$, $X, Y\in\mathfrak{so}(3)$, by means of which $\mathfrak{so}(3)$ can be identified with the dual ${\mathfrak{so}(3)}^{\ast}$. Note that $\mathcal{J}$ is symmetric with respect to $\langle\cdot, \cdot \rangle$. We assume that $\mathcal{J}$ is positive-definite and set ${\displaystyle H(M)=\frac{1}{2}\langle M, \mathcal{J}^{-1}(M)\rangle}$, $M\in{\mathfrak{so}(3)}^{\ast}$. For this function $H$, the Hamiltonian vector field $X_{\widetilde{H}}$ can be given as $X_{\widetilde{H}}=\left(X_{\widetilde{H}}^{\prime}, X_{\widetilde{H}}^{\prime\prime}\right)=\left({L_g}_{\ast}\mathcal{J}^{-1}(M), \mathrm{ad}_{\mathcal{J}^{-1}(M)}^{\ast}M\right)$. In particular, the motion of a free rigid body can be  described by the following Euler equation posed on the angular momentum: 
\[
\frac{\mathsf{d}}{\mathsf{d}t}M=\left[M, \Omega\right], 
\]
where $\Omega=\mathcal{J}^{-1}(M)$. Equivalently, the Lie algebra isomorphism $R:\left(\mathbb{R}^3,\times\right)\rightarrow\mathfrak{so}(3)$ transforms the Euler equation into 
\[
\frac{\mathsf{d}}{\mathsf{d}t}P=P\times \left(\mathcal{I}^{-1}(P)\right), 
\]
where $P=R^{-1}(M)\in\mathbb{R}^3$, $\Omega=\mathcal{J}^{-1}(M)=R\left(\mathcal{I}^{-1}\left(R^{-1}(M)\right)\right)$ with a suitable symmetric positive-definite linear operator $\mathcal{I}:\mathbb{R}^3\rightarrow\mathbb{R}^3$. If $\mathsf{J}=\mathrm{diag}\left(J_1,J_2,J_3\right)$, then we have a matrix presentation as $\mathcal{I}=\mathrm{diag}\left(I_1,I_2,I_3\right)$, such that $I_1=J_2+J_3$, $I_2=J_3+J_1$, $I_3=J_1+J_2$. In this case, the Euler equation can be described by means of the coordinates $P=\left(p_1,p_2,p_3\right)$ as 
\begin{align}\label{euler}
\frac{\mathsf{d}}{\mathsf{d}t}p_1 &=-\left(\frac{1}{I_2}-\frac{1}{I_3}\right)p_2p_3, \notag \\
\frac{\mathsf{d}}{\mathsf{d}t}p_2 &=-\left(\frac{1}{I_3}-\frac{1}{I_1}\right)p_3p_1, \\
\frac{\mathsf{d}}{\mathsf{d}t}p_3 &=-\left(\frac{1}{I_1}-\frac{1}{I_2}\right)p_1p_2. \notag
\end{align}

The Hamiltonian system $\left(T^{\ast}SO(3), \Theta,\widetilde{H}\right)$ is invariant with respect to the left-action by $SO(3)$ and the momentum mapping $\varphi:T^{\ast}SO(3)\cong SO(3)\times {\mathfrak{so}(3)}^{\ast}\ni(g,M)\mapsto \mathrm{Ad}_{g^{-1}}^{\ast}M\in{\mathfrak{so}(3)}^{\ast}$ is associated with this symmetry. With respect to this $SO(3)$ symmetry, we can apply the Marsden-Weinstein reduction \cite{marsden-weinstein} for the free rigid body dynamics. Denoting the coadjoint orbit which passes through $M\in{\mathfrak{so}(3)}^{\ast}$ by $\mathcal{S}_M$, we have the following commutative diagram: 
\[
\begin{array}{ccc}
\varphi^{-1}(M) & \hookrightarrow & T^{\ast}SO(3) \\
\downarrow     &    & \downarrow \varphi \\
\mathcal{S}_M  & \hookrightarrow  & {\mathfrak{so}(3)}^{\ast}
\end{array}
\]
The quotient of the momentum manifold $\varphi^{-1}(M)$ by the action of the stabilizer $SO(3)_M\subset SO(3)$ at $M\in{\mathfrak{so}(3)}^{\ast}$ is diffeomorphic to $\mathcal{S}_M$ and the reduced symplectic form $\omega_M$ coincides with the orbit symplectic (Kirillov-Kostant-Souriau) form. Further, the reduced Hamiltonian on $\mathcal{S}_M$ is the restriction $H|_{\mathcal{S}_M}$. The reduced system $\left(\mathcal{S}_M, \omega_M, H|_{\mathcal{S}_M}\right)$ is completely integrable in the sense of Liouville, since $\mathrm{dim}\mathcal{S}_M=2$, except the orbit which passes through the origin. 

In fact, the Euler equation can also be described by means of the Lie-Poisson bracket $\left\{\cdot,\cdot\right\}$ on ${\mathfrak{so}(3)}^{\ast}$ defined through 
\[
\left\{F,G\right\}(M):=\left\langle M,\left[\nabla F(M), \nabla G(M)\right]\right\rangle, 
\]
where $M\in{\mathfrak{so}(3)}^{\ast}$, $F,G\in\mathcal{C}^{\infty}\left({\mathfrak{so}(3)}^{\ast}\right)$, and where $\nabla F(M)$ denotes the gradient of $F$ at $M$ with respect to $\left\langle\cdot,\cdot\right\rangle$, i.e. $\mathsf{d}F\cdot N=\left\langle N,\nabla F(M)\right\rangle$, for all $N\in{\mathfrak{so}(3)}^{\ast}$. One can easily verify that $\left({\mathfrak{so}(3)}^{\ast},\left\{\cdot,\cdot\right\}\right)$ is a Poisson manifold. For a function $G\in\mathcal{C}^{\infty}\left({\mathfrak{so}(3)}^{\ast}\right)$, the corresponding Hamiltonian vector field $X_G$ with respect to the Poisson bracket $\langle\cdot,\cdot\rangle$ is given through $X_G(F)=\left\{ G, F\right\}$, $F\in\mathcal{C}^{\infty}\left({\mathfrak{so}(3)}^{\ast}\right)$, which can be expressed as $X_G(M)=\mathrm{ad}_{\mathsf{d}G_M}^{\ast}M$. In particular, the Euler equation for the free rigid body dynamics can be given as the Hamilton's equation for the Hamiltonian ${\displaystyle H(M)=\frac{1}{2}\left\langle M, \mathcal{J}^{-1}(M)\right\rangle}$. Further, the function ${\displaystyle L(M)=\frac{1}{2}\left \langle M, M\right\rangle}$, $M\in{\mathfrak{so}(3)}^{\ast}$, which stands for the squared norm of the angular momentum of the rigid body, is a Casimir function in the sense that $\left\{ L, \cdot\right\}=0$. It is straightforward to see that $L$ is a first integral of the Euler equation. The symplectic leaves of the Poisson manifold $\left({\mathfrak{so}(3)}^{\ast},\left\{\cdot,\cdot\right\}\right)$ are the coadjoint orbits $\mathcal{S}_M$, which are described as  level surfaces of the Casimir function $L$. 

The phase portrait for the reduced system $\left(\mathcal{S}_M,\omega_M, H|_{\mathcal{S}_M}\right)$ is well-known. (See, for example, the front cover of \cite{marsden-ratiu}.)  If we take the orthogonal coordinates of $\mathfrak{so}(3)\cong \mathbb{R}^3$, so that the inertia tensor $\mathcal{I}$, or equivalently $\mathcal{J}$, is diagonal, the equilibria of the Euler equation are sitting on the three axes of the coordinate system. The four stationary points on the two axes corresponding to the long and short principal axes, i.e. the largest and the smallest eigenvalues of $\mathcal{I}$, are stable, while the other two on the axis corresponding to the middle axis of $\mathcal{I}$ are unstable. For example, if $\mathcal{I}=\mathrm{diag}(I_1,I_2,I_3)$ with respect to the coordinate system $P=(p_1,p_2,p_3)$ and if $I_1<I_2<I_3$, then the four intersections of $\mathcal{S}_M$ with the $p_1$- and $p_3$-axes are stable equilibria of the system, and the two intersections of $\mathcal{S}_M$ and the $p_2$-axis are unstable. Note that the coadjoint orbit $\mathcal{S}_M$ is a sphere $p_1^2+p_2^2+p_3^2=\text{const.}$

\section{Birkhoff normal forms for a system of one degree of freedom and period integrals}
In this section, we discuss the relation between the Birkhoff normal forms for a Hamiltonian system of one degree of freedom and period integrals. In fact, the derivative of the inverse for the Birkhoff normal forms can be represented by a period integral. 

We consider a Hamiltonian system given by a Hamiltonian $H$ on two-dimensional phase space equipped with a symplectic form $\omega$. For the systems with one degree of freedom, the convergence of the Birkhoff normal form for an equilibrium was essentially known from the study by Siegel \cite{siegel-moser}. 
Let $(x,y)$ be a Darboux coordinate system such that $\omega=\mathsf{d}x\wedge \mathsf{d}y$. Assume that the origin $(x,y)=(0,0)$ is an elliptic equilibrium, and that the Hamiltonian $H$ can be expressed by a Birkhoff normal form ${\displaystyle H=\mathcal{H}\left(\frac{x^2+y^2}{2}\right)}$, where $\mathcal{H}$ is an invertible analytic function around the origin. 

Assume that there is a one-form $\eta$ defined on $U\setminus\left\{(0,0)\right\}$, where $U$ is a neighbourhood of the origin $(0,0)$, such that $\omega=\eta\wedge \mathsf{d}H$. We denote the inverse of the function $\mathcal{H}$ by $\Phi$. Then, we have the following theorem. 
\begin{theorem}\label{thm3.1}
The derivative of the inverse $\Phi$ of the Birkhoff normal form $\mathcal{H}$ around an elliptic stationary point, where $H=0$, writes 
\[
\Phi^{\prime}\left(h\right)=-\frac{1}{2\pi}\int_{H=h}\eta. 
\]
\end{theorem}
\begin{remark}
The one-form can be replaced by $\eta+f\mathsf{d}H$, where $f$ is a function. In fact, if $\eta^{\prime}$ satisfies $\eta^{\prime}\wedge\mathsf{d}H=\omega$, we have $\left(\eta-\eta^{\prime}\right)\wedge\mathsf{d}H=0$. Thus, we have $\eta-\eta^{\prime}=f\mathsf{d}H$, for a suitable function $f$. See \cite{derham} and \cite[\S 4]{vey} for the general case. 
\end{remark}
\noindent\textbf{Proof of Theorem \ref{thm3.1}.}\; We first show that if $\eta_1\wedge \mathsf{d}H=\eta_2\wedge\mathsf{d}H=\omega$ for two one-forms $\eta_1$, $\eta_2$, we have 
\[
\int_{H=h}\eta_1=\int_{H=h}\eta_2.
\]
This is easy because $\left(\eta_1-\eta_2\right)\wedge\mathsf{d}H=0$ means that $\eta_1-\eta_2=f\mathsf{d}H$ for a suitable function $f$ as in the remark, and ${\displaystyle \int_{H=h}\eta_1-\int_{H=h}\eta_2=\int_{H=h}f\mathsf{d}H=0}$. 

We next take the derivative of $H$: 
\[
\mathsf{d}H=\mathcal{H}^{\prime}\left(\frac{x^2+y^2}{2}\right)\left(x\mathsf{d}x+y\mathsf{d}y\right). 
\]
Then, it is easy to show that the one-form 
\[
\eta=\frac{1}{\mathcal{H}^{\prime}\left(\frac{x^2+y^2}{2}\right)}\left\{(1-s)\frac{\mathsf{d}x}{y}-s\frac{\mathsf{d}y}{x}\right\}
\]
satisfies $\eta\wedge \mathsf{d}H=\omega$. The parameter $s$ does not affect the integral. Putting $\epsilon:=\Phi(h)=\mathcal{H}^{-1}(h)$, we introduce the curve 
\[
x=\sqrt{2\epsilon}\cos\theta,\; y=\sqrt{2\epsilon}\sin\theta,\; \theta: 0\rightarrow 2\pi, 
\]
which coincides with the curve $H=\mathcal{H}\left(\frac{x^2+y^2}{2}\right)=h$ in the phase space. Along this curve, we have 
\[
\eta=\frac{1}{\mathcal{H}^{\prime}(\epsilon)}\left\{(1-s)(-\mathsf{d}\theta)-s\mathsf{d}\theta\right\}=\frac{-1}{\mathcal{H}^{\prime}(\epsilon)}\mathsf{d}\theta, 
\]
so this displays 
\begin{align}\label{proof-thm3.1}
\int_{H=h}\eta=\frac{-1}{\mathcal{H}^{\prime}\left(\mathcal{H}^{-1}(h)\right)}\int_0^{2\pi}\mathsf{d}\theta=\frac{-2\pi}{\mathcal{H}^{\prime}\left(\mathcal{H}^{-1}(h)\right)}. 
\end{align}
$\hfill \blacksquare$

\medskip

In a parallel manner, we consider an expression of the derivative of the inverse $\Phi$ of the Birkhoff normal form around hyperbolic stationary point. In this case, we can express the Hamiltonian $H$ in  the Birkhoff normal form $H=\mathcal{H}(XY)$ with the coordinate system $(X,Y)$ such that the symplectic form is given as $\omega=\mathsf{d}X\wedge\mathsf{d}Y$. We assume that the symplectic form $\omega$ and the Hamiltonian $H$ are defined on the complexified phase space as a complex analytic two-form and as a complex analytic function. We consider the following real closed arc: 
\[
\gamma :\; X=\sqrt{\epsilon}e^{\sqrt{-1}\theta},\; Y=\sqrt{\epsilon}e^{-\sqrt{-1}\theta},\; \theta: 0\rightarrow 2\pi, 
\]
where $\epsilon:=\Phi(h)=\mathcal{H}^{-1}(h)$. This real closed curve is clearly contained in the complex curve $H=\mathcal{H}(XY)=h$, where $(X,Y)$ is regarded as a complex coordinate system. As in the case of elliptic equilibria, we can show the following theorem, taking a one-form $\eta^{\prime}$ such that $\omega=\eta^{\prime}\wedge\mathsf{d}H$. 
\begin{theorem}\label{thm3.2}
The derivative of the inverse $\Phi$ of the Birkhoff normal form $\mathcal{H}$ for the Hamiltonian $H$ around a hyperbolic equilibrium writes 
\[
\Phi^{\prime}(h)=-\frac{\sqrt{-1}}{2\pi}\int_{\gamma}\eta^{\prime}. 
\]
\end{theorem}

\begin{remark}
The proof of this theorem can be performed in the same way as in the elliptic case. The choice of the integral path $\gamma$ for the period integral for the hyperbolic stationary point can be explained as follows: \\
We first consider the action-angle coordinates $(x,y)$ around an elliptic stationary point, so that we have $\omega=\mathsf{d}x\wedge\mathsf{d}y$ and ${\displaystyle H=\mathcal{H}\left(\frac{x^2+y^2}{2}\right)}$. Then, we consider the imaginary transformation 
\[
X=\frac{x+\sqrt{-1}y}{\sqrt{2}},\quad Y=\frac{x-\sqrt{-1}y}{\sqrt{2}}. 
\]
The Hamiltonian is rewritten as $H=\mathcal{H}(XY)$ and the origin $(X,Y)=(0,0)$ can be considered as a hyperbolic stationary point with $(X,Y)$ being regarded as real coordinates, if the symplectic form is $\mathsf{d}X\wedge\mathsf{d}Y=-\sqrt{-1}\omega$. Thus, in order to calculate the integral for the hyperbolic stationary point, it suffices to replace $\eta$ by $\eta^{\prime}=-\sqrt{-1}\eta$, for which $\eta^{\prime}\wedge\mathsf{d}H=-\sqrt{-1}\eta\wedge\mathsf{d}H=-\sqrt{-1}\omega=\mathsf{d}X\wedge\mathsf{d}Y$. Note that the path 
$x=\sqrt{2\epsilon}\cos\theta$, $y=\sqrt{2\epsilon}\sin\theta$ is mapped to $\gamma: X=\sqrt{\epsilon}e^{\sqrt{-1}\theta}$, $Y=\sqrt{\epsilon}e^{-\sqrt{-1}\theta}$. By \eqref{proof-thm3.1} in the proof of the previous theorem, we have 
\[
\int_{\gamma}\eta^{\prime}=\frac{2\pi\sqrt{-1}}{\mathcal{H}^{\prime}\left(\mathcal{H}^{-1}(h)\right)}, 
\]
which implies the theorem. 
\end{remark}

Here, we make a comment about the above theorems from the view point of complex analytic geometry. Assume that the symplectic form $\omega$ and the Hamiltonian function $H$ are defined on the complexified phase space $U$ as before. Then, the function $H$ gives rise to a complex analytic fibration of $U$ to a small open set $N$ of the complex line: $\pi: U\rightarrow N$. We assume that $N$ is a neighbourhood of the origin and that the origin is an isolated non-degenerate critical value of $\pi$ in the sense of Morse. In other words, the singular point in the fibre $\pi^{-1}(0)$ is of type $A_1$. This is equivalent to say that the Hamiltonian vector field for $H$ with respect to the symplectic form $\omega$ has non-degenerate critical point, when $H=0$. Moreover, the fibration $\pi:U\rightarrow N$ can be seen as a deformation of $A_1$-singularity. 

Clearly, the level set $H=h$ of the Hamiltonian is a complex analytic curve in $U$. The real integral curve around the elliptic stationary point, which we denote by $\gamma_0$, and the real closed arc $\gamma$ around the hyperbolic stationary point are both vanishing cycles with respect to the deformation $\pi: U\rightarrow N$. If there is only one singular point on $\pi^{-1}(0)$, there exists only one vanishing cycle in the first homology group of the regular fibres of $\pi$ around the singular fibre $\pi^{-1}(0)$. 

In view of the descriptions of the integral paths, the results  of the previous two theorems can be regarded as an expression of the derivative of the inverse of the Birkhoff normal form by a period integral along a vanishing cycle. In the the next section, we apply these two theorems for the free rigid body dynamics. 

\section{Derivative of inverse Birkhoff normal forms for free rigid body dynamics}
In this section, we apply the results in the previous section to the reduced system of the free rigid body dynamics onto the level surfaces of the first integral $L$. We perform the calculation in the coordinates $P=
(p_1,p_2,p_3)$. Assume that the inertia tensor is diagonal: $\mathcal{I}=\mathrm{diag}(I_1,I_2,I_3)$ and $I_1<I_2<I_3$. On the level surface ${\displaystyle L=\frac{1}{2}\left(p_1^2+p_2^2+p_3^2\right)=l}$, where $l$ is a positive constant, there are six equilibria on the $p_1$-, $p_2$-, $p_3$-axes, and those four on the $p_1$- and $p_3$-axes are elliptic, while the other two on the $p_2$-axis are hyperbolic. The reduced Hamiltonian on $L(P)=l$ is the restriction ${\displaystyle H|_{L=l}=\frac{1}{2}\left(\frac{p_1^2}{I_1}+\frac{p_2^2}{I_2}+\frac{p_3^2}{I_3}\right)}$ and the orbit symplectic form $\omega$ on $L(P)=l$ is given as 
\begin{equation}\label{orbit-symplectic-form}
\omega=\frac{\mathsf{d}p_1\wedge\mathsf{d}p_2}{3p_3}=\frac{\mathsf{d}p_2\wedge\mathsf{d}p_3}{3p_1}=\frac{\mathsf{d}p_3\wedge\mathsf{d}p_1}{3p_2}. 
\end{equation}
To see this, we consider the Poison bracket $\left\{\cdot,\cdot\right\}$. For smooth functions $f$ and $g$, we have 
\[
\left\{f,g\right\}(P)=p_1\left(\frac{\partial f}{\partial p_2}\frac{\partial g}{\partial p_3}-\frac{\partial f}{\partial p_3}\frac{\partial g}{\partial p_2}\right)+p_2\left(\frac{\partial f}{\partial p_3}\frac{\partial g}{\partial p_1}-\frac{\partial f}{\partial p_1}\frac{\partial g}{\partial p_3}\right)+p_3\left(\frac{\partial f}{\partial p_1}\frac{\partial g}{\partial p_2}-\frac{\partial f}{\partial p_2}\frac{\partial g}{\partial p_1}\right). 
\]
On the orbit, we have $p_1\mathsf{d}p_1+p_2\mathsf{d}p_2+p_3\mathsf{d}p_3=0$, so that the Poisson bracket restricted to the orbit is given as 
\[
\left\{f,g\right\}|_{L=l}(P)=3p_3\left(f_{p_1}g_{p_2}-f_{p_2}g_{p_1}\right)=3p_3\left(f_{p_2}g_{p_3}-f_{p_3}g_{p_2}\right)=3p_2\left(f_{p_3}g_{p_1}-f_{p_1}g_{p_3}\right). 
\]
The expression of the orbit symplectic form $\omega$ follows from this. 

We consider the derivative of the inverse Birkhoff normal form around the elliptic equilibrium $(p_1,p_2,p_3)=(\sqrt{2l}, 0,0)$, where $(p_2,p_3)$ serves as the local coordinate system on the reduced space. On this coordinate neighbourhood, we consider the one-form 
\begin{equation}\label{associate-system}
\eta_s:=(1-s)\frac{\mathsf{d}p_2}{3\left(\frac{1}{I_3}-\frac{1}{I_1}\right)p_3p_1}+s\frac{\mathsf{d}p_3}{3\left(\frac{1}{I_1}-\frac{1}{I_2}\right)p_1p_2}, 
\end{equation}
where $s$ is an arbitrary parameter. It is easy to verify that $\eta_s\wedge \mathsf{d}H=\omega$ for any $s$. 

Denote the inverse function of the Birkhoff normal form ${\displaystyle H-\frac{l}{2I_1}=\mathcal{H}_1}$ around the elliptic stationary point $(\sqrt{2l}, 0,0)$ by $\Phi_1$. We can derive an expression of it in terms of a period integral as follows: 
\begin{theorem}\label{thm4.1}
The derivative of the inverse Birkhoff normal form around $(p_1, p_2,p_3)=(\pm\sqrt{2l}, 0, 0)$ writes 
\begin{align}\label{dibnf-p_1axis}
\Phi_1^{\prime}(h)=-\frac{1}{3\pi}\sqrt{\frac{2}{l}}\frac{1}{\sqrt{(d-c)(a-b)}}\mathcal{K}\left(\frac{(d-a)(b-c)}{(d-c)(b-a)}\right), 
\end{align}
where ${\displaystyle a=\frac{1}{I_1}, b=\frac{1}{I_2}, c=\frac{1}{I_3}, d=\frac{h}{l}}$ and where $\mathcal{K}$ is the complete elliptic integral of the first kind 
\[
\mathcal{K}(\lambda):=\int_0^1\frac{\mathsf{d}x}{\sqrt{(1-x^2)(1-\lambda x^2)}}. 
\]
\end{theorem}
Denote the right hand side of \eqref{dibnf-p_1axis} by $S(a,b,c,d)$. 

\noindent\textbf{Proof.}\; By Theorem \ref{thm3.1}, we have the expression ${\displaystyle \Phi_1^{\prime}(h)=-\frac{1}{2\pi}\int_{H=h}\eta_s}$. The integral path is given by 
\begin{align}
\left\{
\begin{array}{l}
p_1^2+p_2^2+p_3^2=2l, \\
\frac{p_1^2}{I_1}+\frac{p_2^2}{I_2}+\frac{p_3^2}{I_3}=2h. 
\end{array} \right. \notag
\end{align}
Using these equations, we have 
\[
\int_{H=h}\eta_s=2\left\{(1-s)\int_{-\sqrt{2l\frac{\frac{1}{I_1}-\frac{h}{l}}{\frac{1}{I_1}-\frac{1}{I_2}}}}^{\sqrt{2l\frac{\frac{1}{I_1}-\frac{h}{l}}{\frac{1}{I_1}-\frac{1}{I_2}}}}\frac{\mathsf{d}p_2}{3\left(\frac{1}{I_3}-\frac{1}{I_1}\right)p_3p_1}+s\int_{-\sqrt{2l\frac{\frac{1}{I_1}-\frac{h}{l}}{\frac{1}{I_1}-\frac{1}{I_3}}}}^{\sqrt{2l\frac{\frac{1}{I_1}-\frac{h}{l}}{\frac{1}{I_1}-\frac{1}{I_3}}}}\frac{\mathsf{d}p_3}{3\left(\frac{1}{I_1}-\frac{1}{I_2}\right)p_1p_2}\right\}. 
\]
In the right hand side, the first term can be calculated as follows: 
\begin{align}\label{integral_1}
\int_{-\sqrt{2l\frac{\frac{1}{I_1}-\frac{h}{l}}{\frac{1}{I_1}-\frac{1}{I_2}}}}^{\sqrt{2l\frac{\frac{1}{I_1}-\frac{h}{l}}{\frac{1}{I_1}-\frac{1}{I_2}}}}&\frac{\mathsf{d}p_2}{3\left(\frac{1}{I_3}-\frac{1}{I_1}\right)p_3p_1} \notag \\
=&\int_{-\sqrt{2l\frac{\frac{1}{I_1}-\frac{h}{l}}{\frac{1}{I_1}-\frac{1}{I_2}}}}^{\sqrt{2l\frac{\frac{1}{I_1}-\frac{h}{l}}{\frac{1}{I_1}-\frac{1}{I_2}}}}\frac{\mathsf{d}p_2}{3\sqrt{\left\{2l\left(\frac{h}{l}-\frac{1}{I_1}\right)+\left(\frac{1}{I_1}-\frac{1}{I_2}\right)p_2^2\right\}\left\{2l\left(\frac{1}{I_3}-\frac{h}{l}\right)+\left(\frac{1}{I_2}-\frac{1}{I_3}\right)p_2^2\right\}}} \notag \\
=&\frac{1}{6l\sqrt{\left(\frac{h}{l}-\frac{1}{I_1}\right)\left(\frac{1}{I_3}-\frac{h}{l}\right)}}\int_{-\sqrt{2l\frac{\frac{1}{I_1}-\frac{h}{l}}{\frac{1}{I_1}-\frac{1}{I_2}}}}^{\sqrt{2l\frac{\frac{1}{I_1}-\frac{h}{l}}{\frac{1}{I_1}-\frac{1}{I_2}}}}\frac{\mathsf{d}p_2}{\sqrt{\left(1-\frac{\frac{1}{I_1}-\frac{1}{I_2}}{\frac{1}{I_1}-\frac{h}{l}}\frac{p_2^2}{2l}\right)\left(1-\frac{\frac{1}{I_3}-\frac{1}{I_2}}{\frac{1}{I_3}-\frac{h}{l}}\frac{p_2^2}{2l}\right)}} \notag \\
=&\frac{1}{3\sqrt{2l}\sqrt{\left(\frac{h}{l}-\frac{1}{I_3}\right)\left(\frac{1}{I_1}-\frac{1}{I_2}\right)}}\int_{-1}^1\frac{\mathsf{d}x}{\sqrt{(1-x^2)\left\{1-\frac{\left(\frac{h}{l}-\frac{1}{I_1}\right)\left(\frac{1}{I_2}-\frac{1}{I_3}\right)}{\left(\frac{h}{l}-\frac{1}{I_3}\right)\left(\frac{1}{I_2}-\frac{1}{I_1}\right)}x^2\right\}}}, \notag \\
=& \sqrt{\frac{2}{l}}\frac{1}{3\sqrt{(d-c)(a-b)}}\mathcal{K}\left(\frac{(d-a)(b-c)}{(d-c)(b-a)}\right), 
\end{align}
where we set ${\displaystyle x=p_2\sqrt{\frac{\frac{1}{I_1}-\frac{1}{I_2}}{2l\left(\frac{1}{I_1}-\frac{h}{l}\right)}}}$. 
Similarly, the second integral can be calculated as 
\begin{align}\label{integral_2}
\int_{-\sqrt{2l\frac{\frac{1}{I_1}-\frac{h}{l}}{\frac{1}{I_1}-\frac{1}{I_3}}}}^{\sqrt{2l\frac{\frac{1}{I_1}-\frac{h}{l}}{\frac{1}{I_1}-\frac{1}{I_3}}}}\frac{\mathsf{d}p_3}{\left(\frac{1}{I_1}-\frac{1}{I_2}\right)p_3p_1}=\sqrt{\frac{2}{l}}\frac{1}{3\sqrt{(d-b)(a-c)}}\mathcal{K}\left(\frac{(d-a)(c-b)}{(d-b)(c-a)}\right). 
\end{align}
Note that, setting ${\displaystyle \lambda=\frac{(d-a)(c-b)}{(d-b)(c-a)}}$, we have ${\displaystyle 1-\lambda=\frac{(d-c)(a-b)}{(d-b)(a-c)}}$ and ${\displaystyle \frac{\lambda}{\lambda-1}=\frac{(d-a)(b-c)}{(d-c)(b-a)}}$. Since $\lambda<0$ near the elliptic stationary point $(\sqrt{2l}, 0, 0)$ on the $p_1$-axis, we can use the formula (8.128.1) in \cite{gradshteyn-ryzhik} which displays 
\begin{align}\label{modular-function}
\mathcal{K}\left(\frac{\lambda}{\lambda-1}\right)=\sqrt{1-\lambda}\mathcal{K}(\lambda), \quad \text{for}\; \mathrm{Im}\sqrt{\lambda}<0. 
\end{align}
Thus, the above two integrals \eqref{integral_1} and \eqref{integral_2} are equal to each other. Now, the formula \eqref{dibnf-p_1axis} follows immediately. $\hfill \blacksquare$

Here, we mention some  covariant properties of the Euler equation \eqref{euler}. First, obviously, the transformation 
\[
\delta_1:\begin{bmatrix}p_1 \\ p_2 \\ p_3\end{bmatrix}\mapsto \begin{bmatrix}-p_1 \\ p_2 \\ p_3\end{bmatrix}, \qquad t\mapsto -t
\]
preserves the Euler equation, as well as 
\[
\delta_2:\begin{bmatrix}p_1 \\ p_2 \\ p_3\end{bmatrix}\mapsto \begin{bmatrix}p_1 \\ -p_2 \\ p_3\end{bmatrix}, \qquad t\mapsto -t \qquad \text{and} \qquad 
\delta_3:\begin{bmatrix}p_1 \\ p_2 \\ p_3\end{bmatrix}\mapsto \begin{bmatrix}p_1 \\ p_2 \\ -p_3\end{bmatrix}, \qquad t\mapsto -t. 
\]
It is clear that $\delta_1$, $\delta_2$, $\delta_3$ generate a group isomorphic to $\mathbb{Z}_2\times\mathbb{Z}_2\times\mathbb{Z}_2$. On the other hand, the transformations 
\[
\epsilon_1:\begin{bmatrix}p_1 \\ p_2 \\p_3\end{bmatrix}\mapsto -\begin{bmatrix}p_1 \\ p_3 \\p_2\end{bmatrix}, \qquad \begin{bmatrix}I_1 \\ I_2 \\ I_3\end{bmatrix}\mapsto\begin{bmatrix}I_1 \\ I_3 \\ I_2\end{bmatrix}, 
\]
\[
\epsilon_2:\begin{bmatrix}p_1 \\ p_2 \\p_3\end{bmatrix}\mapsto -\begin{bmatrix}p_3 \\ p_2 \\p_1\end{bmatrix}, \qquad \begin{bmatrix}I_1 \\ I_2 \\ I_3\end{bmatrix}\mapsto\begin{bmatrix}I_3 \\ I_2 \\ I_1\end{bmatrix}, 
\]
\[
\epsilon_3:\begin{bmatrix}p_1 \\ p_2 \\p_3\end{bmatrix}\mapsto -\begin{bmatrix}p_2 \\ p_1 \\p_3\end{bmatrix}, \qquad \begin{bmatrix}I_1 \\ I_2 \\ I_3\end{bmatrix}\mapsto\begin{bmatrix}I_2 \\ I_1 \\ I_3\end{bmatrix}, 
\]
where $t$ is not transformed, preserves the Euler equation. These involutions $\epsilon_1$, $\epsilon_2$, $\epsilon_3$ generate another group isomorphic to the symmetric group $\mathfrak{S}_3$ of degree three. Needless to say that the first integral $L$ is invariant with respect to the  above transformations. Note that the transformations $\delta_1$, $\delta_2$, $\delta_3$, $\epsilon_1$, $\epsilon_2$, $\epsilon_3$ act on the symplectic form $\omega$ and the Hamiltonian $H$ as $\delta_j^{\ast}\omega=-\omega$, $j=1,2,3$, $\delta_j^{\ast}H=H$, $j=1,2,3$; $\epsilon_j^{\ast}\omega=-\omega$ and $\epsilon_j^{\ast}H=H$, $j=1,2,3$. 

By the transformation $\delta_1$ of the Euler equation, the equilibrium $(p_1,p_2,p_3)=(\sqrt{2l}, 0,0)$ is mapped to $(-\sqrt{2l}, 0,0)$. The integral curves around the two equilibrium points $(\pm\sqrt{2l},0,0)$ are mapped to each other, but their orientations are opposite, since the time is reversed by $\delta_1$. As to the period integral, the integrand $\eta_s$ is transformed as $\delta_1^{\ast}\eta_s=-\eta_s$. Thus, the integral ${\displaystyle \int_{H=h}\eta_s}$ is invariant with respect to $\delta_1$. As a result, we have the following corollary. 
\begin{corollary}\label{cor4.2}
The derivative of the inverse of the Birkhoff normal form around $(p_1, p_2,p_3)=(-\sqrt{2l}, 0, 0)$ is given by \eqref{dibnf-p_1axis}. 
\end{corollary}

Similarly, the transformation $\epsilon_2$ maps the equilibrium $(p_1,p_2,p_3)=(\sqrt{2l},0,0)$ to $(0,0,-\sqrt{2l})$. It maps the integral curves around each equilibrium to each other by $\epsilon_2$, reversing their orientations. The integrand $\eta_s$ of the period integral is transformed as $\epsilon_2^{\ast}\eta_s=-\eta_s$ and the period integral itself is transformed from \eqref{dibnf-p_1axis} by the permutation $(ac)\in\mathfrak{S}_4$ with respect to $\epsilon_2$. 
\begin{corollary}
The derivative of the inverse for the Birkhoff normal form $\Phi_3$ around $(p_1,p_2,p_3)=(0,0,\pm\sqrt{2l})$ writes 
\begin{align}
\Phi_3^{\prime}(h)=-\frac{1}{3\pi}\sqrt{\frac{2}{l}}\frac{1}{\sqrt{(d-a)(c-b)}}\mathcal{K}\left(\frac{(d-c)(b-a)}{(d-a)(b-c)}\right)=S(c,b,a,d), \notag
\end{align}
where ${\displaystyle a=\frac{1}{I_1}, b=\frac{1}{I_2}, c=\frac{1}{I_3}, d=\frac{h}{l}}$ and $\mathcal{K}$ is the complete elliptic integral of the first kind. 
\end{corollary}

Next, we consider the derivative of the inverse for the Birkhoff normal form around the hyperbolic stationary points $(p_1,p_2,p_3)=(0,\pm\sqrt{2l},0)$. Around the stationary point $(p_1,p_2,p_3)=(0,\sqrt{2l},0)$, we can use $(p_3,p_1)$ as a local coordinate system. Take the one-form 
\[
\eta_s^{\prime}:=(1-s)\frac{\mathsf{d}p_3}{3\left(\frac{1}{I_1}-\frac{1}{I_2}\right)p_1p_2}+s\frac{\mathsf{d}p_1}{3\left(\frac{1}{I_2}-\frac{1}{I_3}\right)p_2p_3}, 
\]
where $s$ is an arbitrary parameter as before, and the real closed arc 
\[
\gamma:\; p_3=\sqrt{2l\frac{\frac{1}{I_2}-\frac{h}{l}}{\frac{1}{I_2}-\frac{1}{I_3}}}\cos\theta, \quad p_1=\sqrt{2l\frac{\frac{1}{I_2}-\frac{h}{l}}{\frac{1}{I_2}-\frac{1}{I_1}}}\sin\theta, 
\]
with the parameter $\theta:0\rightarrow 2\pi$. Note that $\gamma$ is not the real integral curve, since either $p_3$ or $p_1$ is purely imaginary in general, although it is a closed real curve contained in the complex solution of the system. The complex solution is given by the complexified affine arc 
\begin{align}
p_1^2+p_2^2+p_3^2 &= 2l, \notag \\
\frac{p_1^2}{I_1}+\frac{p_2^2}{I_2}+\frac{p_3^2}{I_3} &=2h, \notag
\end{align}
where $(p_1,p_2,p_3)\in\mathbb{C}^3$ are regarded as complex affine coordinates. 

We consider the period integral ${\displaystyle \int_{\gamma}\eta_s^{\prime}}$. Note that the equilibrium $(p_1,p_2,p_3)=(\sqrt{2l},0,0)$ is mapped to $(0,\sqrt{2l},0)$ by the transformation $\epsilon_3\circ\delta_1$, which maps the integral curves around $(\sqrt{2l},0,0)$ to those represented by $\gamma$. Therefore, using Theorem \ref{thm3.2}, we can calculate the derivative for the inverse of the Birkhoff normal form around $(0,\sqrt{2l},0)$ as 
\[
\Phi_2^{\prime}(h)=\frac{\sqrt{-1}}{3\pi}\sqrt{\frac{2}{l}}\frac{1}{\sqrt{(d-c)(b-a)}}\mathcal{K}\left(\frac{(d-b)(a-c)}{(d-c)(a-b)}\right)=-\sqrt{-1}S(b,a,c,d). 
\]
As before, this result is invariant with respect to the transformation $\delta_2$, which maps $(p_1,p_2,p_3)=(0,\sqrt{2l},0)$ to $(0,-\sqrt{2l},0)$. 
\begin{theorem}
The derivative of the inverse $\Phi_2$ of the Birkhoff normal forms around $(p_1,p_2,p_3)=(0,\pm\sqrt{2l},0)$ writes 
\begin{align}
\Phi_2^{\prime}(h)=\frac{\sqrt{-1}}{3\pi}\sqrt{\frac{2}{l}}\frac{1}{\sqrt{(d-c)(b-a)}}\mathcal{K}\left(\frac{(d-b)(a-c)}{(d-c)(a-b)}\right)=-\sqrt{-1}S(b,a,c,d),  
\end{align}
where ${\displaystyle a=\frac{1}{I_1}, b=\frac{1}{I_2}, c=\frac{1}{I_3}, d=\frac{h}{l}}$ and $\mathcal{K}$ is the complete elliptic integral of the first kind. 
\end{theorem}

Before closing this section, we give the following formulae for $S(a,b,c,d)$ obtained through the transformations $\delta_1, \delta_2, \delta_3, \epsilon_1, \epsilon_2, \epsilon_3$: 
\begin{align}\label{s_3-formulae}
S(a,b,c,d)=S(a,c,b,d), \notag \\
S(b,a,c,d)=S(b,c,a,d), \\ 
S(c,b,a,d)=S(c,a,b,d). \notag
\end{align}
These three formulae follow from \eqref{modular-function} in the proofs of the previous theorems and the covariance with respect to $\delta_j$, $\epsilon_j$, $j=1,2,3$. The list can be regarded as the description of the action of the symmetric group $\mathfrak{S}_3$ of degree three on $S(a,b,c,d)$. We explain the action by the symmetric group $\mathfrak{S}_4$ of degree four in Section 6. 

\medskip
\medskip

\section{Elliptic fibration and explicit calculation of the cocycles} 
In this section, we discuss the naive elliptic fibration considered in \cite{naruki-tarama}  and calculate the cocycles of the period integrals which appeared in the previous section. We start with a basic description of the naive elliptic fibration. As we have seen the integral curve of the Euler equation is given by the intersection of the two quadrics 
\[
\begin{cases}
p_1^2+p_2^2+p_3^2=2l, \\
\frac{p_1^2}{I_1}+\frac{p_2^2}{I_2}+\frac{p_3^2}{I_3}=2h, 
\end{cases}
\]
where $P=(p_1,p_2,p_3)$ is the angular momentum vector. In the previous section, we have seen that it is useful to think about its complexification, regarding $(p_1,p_2,p_3)$ as affine coordinates in $\mathbb{C}^3$. From the viewpoint of algebraic or analytic geometry, it is further natural to compactify the integral curve by the complex projective curve 
\begin{align}\label{projective_integral_curve}
\begin{cases}
x^2+y^2+z^2+w^2=0, \\
ax^2+by^2+cz^2+dw^2=0,
\end{cases}
\end{align}
where $(x:y:z:w)\in P_3(\mathbb{C})$ are the coordinates given by ${\displaystyle p_1=\sqrt{-2l}\frac{x}{w}, p_2=\sqrt{-2l}\frac{y}{w}, p_3=\sqrt{-2l}\frac{z}{w}}$, and where $a,b,c,d\in\mathbb{C}$ are parameters given by ${\displaystyle \frac{1}{I_1}=a, \frac{1}{I_2}=b, \frac{1}{I_3}=c, \frac{h}{l}=d}$. The following proposition is fundamental to the geometric understanding of this projective curve. 
\begin{proposition}
If $a,b,c,d$ are distinct, then the variety $C$ defined by the above equations \eqref{projective_integral_curve} is a smooth elliptic curve, which has four branch points $a,b,c,d$ as a double covering of the projective line $P_1(\mathbb{C})\cong\mathbb{C}\cup\{\infty\}$. 
\end{proposition}

For the proof, see \cite{naruki-tarama}. Because of this proposition, we have a natural elliptic fibration. In fact, denoting by $F$ the algebraic variety in $P_3(\mathbb{C})\times P_3(\mathbb{C}): \left((x:y:z:w),(a:b:c:d)\right)$ defined by the equation \eqref{projective_integral_curve}, we consider the projection $\pi_F: F\ni  \left((x:y:z:w),(a:b:c:d)\right)\mapsto (a:b:c:d)\in P_3(\mathbb{C})$ to the second component of the product space, which is an elliptic fibration. The fibration $\pi_F:F\rightarrow P_3(\mathbb{C})$ is called the naive elliptic fibration. Here, an elliptic fibration means a smooth holomorphic mapping of a complex space onto another complex space whose regular fibres are elliptic curves. As basic facts of the naive elliptic fibration $\pi_F:F\rightarrow P_3(\mathbb{C})$, it is known that the total space $F$ is smooth rational variety and that the fibration $\pi_F$ is non-flat, i.e. there is a two-dimensional fibre of $\pi_F$. In fact, the singular fibres of $\pi_F$ are classified in \cite{naruki-tarama} as follows: 

\medskip

\textbf{Classification of the singular fibres of $\pi_F$}
\begin{enumerate}
\item If only two of the parameters $a,b,c,d$ are equal, the fibre consists of two smooth rational curves intersecting at two points. This is a singular fibre of type $\mathrm{I}_2$ in Kodaira's notation \cite{kodaira, barth-hulek-peters-vandeven}. 
\item If two of $a,b,c,d$ are equal and the other two are also equal without further coincidence, the fibre consists of four smooth rational curves intersecting cyclically. This is a singular fibre of type $\mathrm{I}_4$ in Kodaira's notation. 
\item If three of $a,b,c,d$ are equal without further coincidence, the fibre is a smooth rational curve, as a point set, but with multiplicity two. This singular fibre is not in the list of singular fibres of elliptic surfaces by Kodaira. 
\item If $a=b=c=d$, the fibre is a space quadric surface $x^2+y^2+z^2+w^2=0$. 
\end{enumerate}
It is also known that the fibration $\pi_F$ has no (meromorphic) section. On the total space $F$, there is an action of the group $\mathbb{Z}_2\times \mathbb{Z}_2$ which respects the fibration $\pi_F$ and which has no fixed point on the regular fibres. Taking the quotient, we have an elliptic fibration bimeromorphic to a Weierstra{\ss} normal form, which is regular and flat and which admits only singular fibres included in the Kodaira's list, after suitable modifications. See \cite{naruki-tarama} for the detailed discussion. 

\medskip

In order to think about the monodromy of the naive elliptic fibration $\pi_F:F\rightarrow P_3(\mathbb{C})$, we introduce the following locally constant sheaf $G$ on the base space $P_3(\mathbb{C})$. As is known from the description of the singular fibres, the singular locus of the elliptic fibration $\pi_F :F\rightarrow P_3(\mathbb{C})$ is given by the divisor $D:\{a=b\}+ \{a=c\}+ \{a=d\}+ \{b=c\}+ \{b=d\}+ \{c=d\}$ on $P_3(\mathbb{C}):(a:b:c:d)$. For $(a:b:c:d)\in P_3(\mathbb{C})\setminus \mathrm{Supp}(D)$ , the fibre $\pi^{-1}(a:b:c:d)$ is a smooth complex torus, so that the first homology group $H_1\left(\pi_F^{-1}(a:b:c:d); \mathbb{Z}\right)$ with the coefficients in $\mathbb{Z}$ form a locally constant sheaf of $\mathbb{Z}_{P_3(\mathbb{C})\setminus \mathrm{Supp}(D)}$-module 
\[
G^{\prime}=\bigsqcup_{(a:b:c:d)\in P_3(\mathbb{C})\setminus \mathrm{Supp}(D)} H_1 \left(\pi_F^{-1}(a:b:c:d); \mathbb{Z}\right)
\]
over $P_3(\mathbb{C})\setminus \mathrm{Supp}(D)$. For a point $(a:b:c:d)\in \mathrm{Supp}(D)$, we consider its sufficiently small polydisc neighbourhood $U$ and set $U^{\prime}:=U\setminus \mathrm{Supp}(D)$. The group $\Gamma(U^{\prime}, G^{\prime})$ of sections of $G^{\prime}$ is determined independently from the choice of $U$. Regarding $\Gamma(U^{\prime},G^{\prime})$ as the stalk $G_{(a:b:c:d)}$ on $(a:b:c:d)\in\mathrm{Supp}(D)$, we consider the sheaf 
\[
G=G^{\prime}\sqcup\bigsqcup_{(a:b:c:d)\in\mathrm{Supp}(D)}G_{(a:b:c:d)}
\]
over the base space $P_3(\mathbb{C})$. Clearly, it is a locally constant sheaf of $\mathbb{Z}_{P_3(\mathbb{C})}$-module. We call $G$ the homological invariant of $\pi_F$. Taking the dual of each stalk, we have the locally constant sheaf of $\mathbb{Z}_{P_3(\mathbb{C})}$-module 
\[
G^{\ast}={G^{\ast}}^{\prime}\sqcup\bigsqcup_{(a:b:c:d)\in\mathrm{Supp}(D)}G_{(a:b:c:d)}^{\ast}, 
\]
where 
\[
{G^{\ast}}^{\prime}=\bigsqcup_{(a:b:c:d)\in P_3(\mathbb{C})\setminus \mathrm{Supp}(D)} H^1 \left(\pi_F^{-1}(a:b:c:d); \mathbb{Z}\right)
\]
and $G_{(a:b:c:d)}^{\ast}$ for $(a:b:c:d)\in\mathrm{Supp}(D)$ is determined in the same manner as $G_{(a:b:c:d)}$. We call $G^{\ast}$ the cohomological invariant of $\pi_F$. 

Let $p_0$ be a point in $P_3(\mathbb{C})\setminus\mathrm{Supp}(D)$ and $\gamma:t\mapsto \gamma(t)$, $0\leq t\leq 1$, a closed path with the reference point at $p_0$ in $P_3(\mathbb{C})\setminus\mathrm{Supp}(D)$. Taking a basis $\sigma_{1,0}, \sigma_{2,0}$ of $H_1(\pi_F^{-1}(p_0),\mathbb{Z})$, we consider the elements $\sigma_1(t), \sigma_2(t)$ of $H_1(\pi_F^{-1}(\gamma(t)),\mathbb{Z})$ which continuously depend on $t$ and $\sigma_i(0)=\sigma_{i,1}$, $i=1,2$. Then, there is a matrix $A_{[\gamma]}\in SL(2,\mathbb{Z})$ such that 
\[
\begin{bmatrix}
\sigma_1(1) \\ \sigma_2(1)
\end{bmatrix}
=A_{[\gamma]}
\begin{bmatrix}
\sigma_{1,0} \\ \sigma_{2,0}
\end{bmatrix}.
\]
Note that $A_{[\gamma]}$ depends only on the homotopy class $[\gamma]\in \pi_1\left(P_3(\mathbb{C})\setminus \mathrm{Supp}(D); p_0\right)$ of $\gamma$. The homomorphism $\rho: \pi_1\left(P_3(\mathbb{C})\setminus \mathrm{Supp}(D); p_0\right)\ni[\gamma]\mapsto A_{[\gamma]}\in SL(2,\mathbb{Z})$ is called the monodromy representation of $\pi_1\left(P_3(\mathbb{C})\setminus \mathrm{Supp}(D); p_0\right)$. Similarly, taking the dual basis $\sigma_{1,0}^{\ast}, \sigma_{2,0}^{\ast}$ of $H^1(\pi_F^{-1}(p_0), \mathbb{Z})$ such that $\sigma_{i,0}^{\ast}\cdot\sigma_{j,0}=\delta_{ij}$, we have the representation $\rho^{\ast}: \pi_1\left(P_3(\mathbb{C})\setminus \mathrm{Supp}(D); p_0\right)\ni[\gamma]\mapsto A_{[\gamma]}^{\mathrm{T}}\in SL(2,\mathbb{Z})$. We call it also monodromy representation. It is clear that the homological invariant $G$, as well as the cohomological invariant $G^{\ast}$, is determined by the monodromy representation $\rho$ or $\rho^{\ast}$. 

\medskip

We, now, consider the period integrals which appeared in the last section from the viewpoint of the (co)homological invariant $G$ (or $G^{\ast}$) and discuss the global monodromy, namely the monodromy representation itself. We start with the extension of the one-form $\eta$ such that $\omega=\eta\wedge \mathsf{d}H$ to the complex elliptic fibration $\pi_F: F\rightarrow P_3(\mathbb{C})$. On the adjoint orbit ${\displaystyle L(P)=\frac{1}{2}\left(p_1^2+p_2^2+p_3^2\right)=l}$, the orbit symplectic form $\omega$ writes as \eqref{orbit-symplectic-form}. On the coordinate neighbourhood with the coordinates $(p_2,p_3)$, the one-form $\eta_s^{(1)}:=\eta_s$ in \eqref{associate-system} satisfies $\omega=\eta_s^{(1)}\wedge \mathsf{d}H$. Similarly, in the coordinates $(p_3,p_1)$, the one-form 
\[
\eta_s^{(2)}:=(1-s)\frac{\mathsf{d}p_3}{3\left(\frac{1}{I_1}-\frac{1}{I_2}\right)p_1p_2}+s\frac{\mathsf{d}p_1}{3\left(\frac{1}{I_2}-\frac{1}{I_3}\right)p_2p_3}
\]
satisfies $\omega=\eta_s^{(2)}\wedge \mathsf{d}H$, and, in the coordinates $(p_1,p_2)$, the one-form 
\[
\eta_s^{(3)}:=(1-s)\frac{\mathsf{d}p_1}{3\left(\frac{1}{I_2}-\frac{1}{I_3}\right)p_2p_3}+s\frac{\mathsf{d}p_2}{3\left(\frac{1}{I_3}-\frac{1}{I_1}\right)p_3p_1}
\]
satisfies $\omega=\eta_s^{(3)}\wedge\mathsf{d}H$. Note that on the integral curve $H=h$, where $h$ is a constant, we have 
\[
\frac{\mathsf{d}p_1}{\left(\frac{1}{I_2}-\frac{1}{I_3}\right)p_2p_3}=\frac{\mathsf{d}p_2}{\left(\frac{1}{I_3}-\frac{1}{I_1}\right)p_3p_1}=\frac{\mathsf{d}p_3}{\left(\frac{1}{I_1}-\frac{1}{I_2}\right)p_1p_2}, 
\]
so that $\eta_s^{(1)}=\eta_{s^{\prime}}^{(2)}=\eta_{s^{\prime\prime}}^{(3)}$ for arbitrary $s, s^{\prime}, s^{\prime\prime}$ and their integrals along the same cycle are equal. With this in mind, we consider the following one-form $\eta$ as a relative differential form with respect to the fibration $\pi_F:F\rightarrow P_3(\mathbb{C})$: 
\[
\eta=\frac{1}{\sqrt{2l}}\frac{w^2\mathsf{d}\left(\frac{y}{w}\right)}{3(c-a)zx}. 
\]
Because of the transformation of the coordinates ${\displaystyle p_1=\sqrt{-2l}\frac{x}{w}, p_2=\sqrt{-2l}\frac{y}{w}, p_3=\sqrt{-2l}\frac{z}{w}}$ and ${\displaystyle a=\frac{1}{I_1}, b=\frac{1}{I_2}, c=\frac{1}{I_3}, d=\frac{h}{l}}$, we can see that $\eta$ and $\eta_s^{(i)}$, $i=1,2,3$, induces the same section in $\Omega_{\pi_F}^1:=\Omega_F^1/\pi_F^{\ast}\Omega_{P_3(\mathbb{C})}^1$. Moreover, we have the following proposition. 
\begin{proposition}
The one-form $\eta$ induces a globally-defined meromorphic section of $\Omega_{\pi_F}^1$, which is holomorphic non-zero one-form on $P_3(\mathbb{C})\setminus \mathrm{Supp}(D)$. 
\end{proposition}
\noindent\textbf{Proof.}\; On the total space $F$ of the elliptic fibration $\pi_F$, we have 
\begin{equation}\label{meromorphic-one-form}
\begin{array}{rc}
\eta=&\frac{1}{\sqrt{-2l}}\frac{w^2\mathsf{d}\left(\frac{x}{w}\right)}{3(b-c)yz}=\frac{1}{\sqrt{-2l}}\frac{w^2\mathsf{d}\left(\frac{y}{w}\right)}{3(c-a)zx}=\frac{1}{\sqrt{-2l}}\frac{w^2\mathsf{d}\left(\frac{z}{w}\right)}{3(a-b)xy} \\
=&\frac{1}{\sqrt{-2l}}\frac{x^2\mathsf{d}\left(\frac{y}{x}\right)}{3(c-d)zw}=\frac{1}{\sqrt{-2l}}\frac{x^2\mathsf{d}\left(\frac{z}{x}\right)}{3(d-b)wy}=\frac{1}{\sqrt{-2l}}\frac{x^2\mathsf{d}\left(\frac{w}{x}\right)}{3(b-c)yz} \\
=&\frac{1}{\sqrt{-2l}}\frac{y^2\mathsf{d}\left(\frac{z}{y}\right)}{3(d-a)wx}=\frac{1}{\sqrt{-2l}}\frac{y^2\mathsf{d}\left(\frac{w}{y}\right)}{3(a-c)xz}=\frac{1}{\sqrt{-2l}}\frac{y^2\mathsf{d}\left(\frac{x}{y}\right)}{3(c-d)zw} \\
=&\frac{1}{\sqrt{-2l}}\frac{z^2\mathsf{d}\left(\frac{w}{z}\right)}{3(a-b)xy}=\frac{1}{\sqrt{-2l}}\frac{z^2\mathsf{d}\left(\frac{x}{z}\right)}{3(b-d)yw}=\frac{1}{\sqrt{-2l}}\frac{z^2\mathsf{d}\left(\frac{y}{z}\right)}{3(d-a)wx}, 
\end{array}
\end{equation}
where $\eta$, as well as other one-forms, is regarded as a (local) meromorphic section of $\Omega_{\pi_F}^1$. Clearly, $\eta$ is holomorphic and non-zero over $P_3(\mathbb{C})\setminus \mathrm{Supp}(D)$ from \eqref{meromorphic-one-form}. $\hfill \blacksquare$

Note that the integral $\int_{\sigma}\eta$ of the one-form $\eta$ over a cycle $\sigma$ included in a fibre of $\pi_F$ depends only on the class of the relative differential one-form to which $\eta$ belongs. Thus, the correspondence 
\[
\int_{\cdot}\eta: G^{\prime}\supset H_1(\pi_F^{-1}(a:b:c:d),\mathbb{Z})\ni\sigma\mapsto \int_{\sigma}\eta\in \mathbb{C}
\]
can be regarded as a linear functional over $G^{\prime}$ with respect to $\mathbb{C}_{P_3(\mathbb{C})\setminus \mathrm{Supp}(D)}$, so that $\int_{\cdot}\eta\in {{G^{\ast}}^{\prime}}^{\mathbb{C}}:= {G^{\ast}}^{\prime}\otimes_{\mathbb{Z}_{P_3(\mathbb{C})\setminus \mathrm{Supp}(D)}} \mathbb{C}_{P_3(\mathbb{C)}\setminus \mathrm{Supp}(D)}$. It is clear that $\int_{\sigma_{1,0}}\eta$ and $\int_{\sigma_{2,0}}\eta$ form a basis of ${{{G^{\ast}}^{\prime}}^{\mathbb{C}}}_{p_0}=H^1(\pi_F^{-1}(p_0),\mathbb{C})$ and it suffices to consider the monodromy of these period integrals in order to calculate the monodromy representation of $\pi_1(P_3(\mathbb{C})\setminus \mathrm{Supp}(D), p_0)$ with respect to the fibration $\pi_F: F\rightarrow P_3(\mathbb{C})$. 

We choose a basis of the first homology group $H_1(\pi_F^{-1}(p_0),\mathbb{Z})$ of the regular fibre near the elliptic and hyperbolic stationary points of the free rigid body dynamics. Assuming the same condition $I_1< I_2 <I_3$ for the inertia tensor $\mathcal{I}=\mathrm{diag}(I_1,I_2,I_3)$ as in Section 4, we start with a regular fibre around the elliptic stationary points on the $p_1$-axis. Note that these two points are in the same intersection of two quadrics $H=h, L=l$, such that ${\displaystyle \frac{h}{l}=\frac{1}{I_1}}$, i.e. $a=d$. For the simplicity, we assume that ${\displaystyle \frac{1}{I_1}>\frac{h}{l}>\frac{1}{I_2}>\frac{1}{I_3}}$. The real integral curves are parameterized as 
\begin{align}
\sigma_1^{(1)}: (p_1,p_2,p_3) =&\left(\sqrt{-\frac{2l\left(\frac{1}{I_2}-\frac{h}{l}\right)}{\frac{1}{I_1}-\frac{1}{I_2}}\left\{1-\frac{\left(\frac{h}{l}-\frac{1}{I_1}\right)\left(\frac{1}{I_3}-\frac{1}{I_2}\right)}{\left(\frac{h}{l}-\frac{1}{I_2}\right)\left(\frac{1}{I_3}-\frac{1}{I_1}\right)}\sin^2\theta\right\}}, \right. \notag \\
&\qquad\qquad\qquad\qquad\qquad\qquad\left. \sqrt{\frac{2l\left(\frac{1}{I_1}-\frac{h}{l}\right)}{\frac{1}{I_1}-\frac{1}{I_2}}}\cos\theta, \sqrt{\frac{2l\left(\frac{1}{I_1}-\frac{h}{l}\right)}{\frac{1}{I_1}-\frac{1}{I_3}}}\sin\theta\right), \notag
\end{align}
where $\theta:0\rightarrow 2\pi$, near $(p_1,p_2,p_3)=(\sqrt{2l}, 0,0)$, and as 
\begin{align}
\sigma_2^{(1)}: (p_1,p_2,p_3) =&\left(-\sqrt{-\frac{2l\left(\frac{1}{I_2}-\frac{h}{l}\right)}{\frac{1}{I_1}-\frac{1}{I_2}}\left\{1-\frac{\left(\frac{h}{l}-\frac{1}{I_1}\right)\left(\frac{1}{I_3}-\frac{1}{I_2}\right)}{\left(\frac{h}{l}-\frac{1}{I_2}\right)\left(\frac{1}{I_3}-\frac{1}{I_1}\right)}\sin^2\theta\right\}}, \right.\notag \\
&\qquad\qquad\qquad\qquad\qquad\qquad\left. \sqrt{\frac{2l\left(\frac{1}{I_1}-\frac{h}{l}\right)}{\frac{1}{I_1}-\frac{1}{I_2}}}\cos\theta, \sqrt{\frac{2l\left(\frac{1}{I_1}-\frac{h}{l}\right)}{\frac{1}{I_1}-\frac{1}{I_3}}}\sin\theta\right), \notag
\end{align}
where $\theta:0\rightarrow 2\pi$, near $(p_1,p_2,p_3)=(-\sqrt{2l}, 0,0)$. If $h$ is near to ${\displaystyle \frac{l}{I_1}}$, i.e. if $d$ is near to $a$, these real closed arcs are vanishing cycles around two different singular points and are included in the same fibre of $\pi_F: F\rightarrow P_3(\mathbb{C})$. By Theorem \ref{thm4.1} and Corollary \ref{cor4.2}, we see that the period integrals of $\eta$ along $\sigma_1^{(1)}$ and $\sigma_2^{(1)}$ are the same, so that 
\[
\int_{\sigma_1^{(1)}}\eta=\int_{\sigma_2^{(1)}}\eta=S(a,b,c,d). 
\]
We take this period integral as one of the basis of the first cohomology group of the regular fibre of $\pi_F :F\rightarrow P_3(\mathbb{C})$ around the singular locus $a=d$. Note that the corresponding singular fibre is of type $\mathrm{I}_2$ (in Kodaira's sense), so there are two singular points of type $A_1$ in the singular fibre. 

To obtain the other basis element, we consider the following real one-dimensional arcs in the same regular fibre. We take the arcs 
\begin{align}
\tau_{\pm}^{(1)}: (p_1,p_2,p_3) =&\left(\pm\sqrt{-\frac{2l\left(\frac{1}{I_2}-\frac{h}{l}\right)}{\frac{1}{I_1}-\frac{1}{I_2}}\left\{1+\frac{\left(\frac{h}{l}-\frac{1}{I_1}\right)\left(\frac{1}{I_3}-\frac{1}{I_2}\right)}{\left(\frac{h}{l}-\frac{1}{I_2}\right)\left(\frac{1}{I_3}-\frac{1}{I_1}\right)}\sinh^2\varphi\right\}}, \right. \notag \\
&\qquad\qquad\qquad\qquad\qquad\qquad\left. \sqrt{\frac{2l\left(\frac{1}{I_1}-\frac{h}{l}\right)}{\frac{1}{I_1}-\frac{1}{I_2}}}\cosh\varphi, \pm\sqrt{-1}\sqrt{\frac{2l\left(\frac{1}{I_1}-\frac{h}{l}\right)}{\frac{1}{I_1}-\frac{1}{I_3}}}\sinh\varphi\right), \notag
\end{align}
where $\varphi$ moves from $-\infty$ to $+\infty$. It is easy to check that these two arcs are included in the same fibre of $\pi_F:F\rightarrow P_3(\mathbb{C})$ near the singular locus $a=d$. We can describe the behavior of these arcs when $\varphi$ approaches the infinity. For each of the arcs $\tau_{\pm}^{(1)}$, we have 
\begin{align}
(x:y:z:w)=&(p_1:p_2:p_3:\sqrt{-2l}) \notag \\
\rightarrow&\left(\pm\sqrt{\frac{2l\left(\frac{h}{l}-\frac{1}{I_1}\right)\left(\frac{1}{I_3}-\frac{1}{I_2}\right)}{\left(\frac{1}{I_1}-\frac{1}{I_2}\right)\left(\frac{1}{I_3}-\frac{1}{I_1}\right)}}: \sqrt{\frac{2l\left(\frac{1}{I_1}-\frac{h}{l}\right)}{\frac{1}{I_1}-\frac{1}{I_2}}}: \mp \sqrt{\frac{2l\left(\frac{1}{I_1}-\frac{h}{l}\right)}{\frac{1}{I_3}-\frac{1}{I_1}}}:0 \right) \notag \\
&= \left(\pm\sqrt{\frac{1}{I_2}-\frac{1}{I_3}}:\sqrt{\frac{1}{I_3}-\frac{1}{I_1}}:\mp\sqrt{\frac{1}{I_1}-\frac{1}{I_2}}:0\right) \notag \\
&= \left(\pm\sqrt{b-c}:\sqrt{c-a}:\mp\sqrt{a-b}:0\right), \notag 
\end{align}
as $\varphi\rightarrow -\infty$, and 
\[
(x:y:z:w)\rightarrow \left(\pm\sqrt{b-c}:\sqrt{c-a}:\pm\sqrt{a-b}:0\right),
\]
as $\varphi\rightarrow +\infty$. Here, we used ${\displaystyle \frac{\sinh\varphi}{\cosh\varphi}\rightarrow \pm 1}$ as $\varphi\rightarrow\pm\infty$. Since $\cosh\varphi \geq 1$ for $\varphi\in\mathbb{R}$, we see that $\tau_{\pm}^{(1)}$ do not intersect with each other. On the other side, the arc $\tau_+^{(1)}$ intersects with $\sigma_1^{(1)}$ at ${\displaystyle (p_1,p_2,p_3)=\left(\sqrt{-\frac{2l\left(\frac{1}{I_2}-\frac{h}{l}\right)}{\frac{1}{I_1}-\frac{1}{I_2}}},\sqrt{\frac{2l\left(\frac{1}{I_1}-\frac{h}{l}\right)}{\frac{1}{I_1}-\frac{1}{I_2}}}, 0\right)}$ and $\tau_-^{(1)}$ intersect with $\sigma_2^{(1)}$ at $(p_1,p_2,p_3)=$\\${\displaystyle\left(-\sqrt{-\frac{2l\left(\frac{1}{I_3}-\frac{h}{l}\right)}{\frac{1}{I_1}-\frac{1}{I_2}}}, \sqrt{\frac{2l\left(\frac{1}{I_1}-\frac{h}{l}\right)}{\frac{1}{I_1}-\frac{1}{I_2}}},0\right)}$. There are no further intersections among these arcs. We also need to take the following arcs: 
\begin{align}
{\tau_{\pm}^{\prime}}^{(1)}:(p_1,&p_2,p_3) =\left(-\sqrt{-1}\sqrt{\frac{2l\left(\frac{1}{I_2}-\frac{h}{l}\right)}{\frac{1}{I_2}-\frac{1}{I_1}}}\sinh\varphi, \right. \notag \\
&\left.  \pm\sqrt{-\frac{2l\left(\frac{1}{I_3}-\frac{h}{l}\right)}{\frac{1}{I_2}-\frac{1}{I_3}}\left\{1+\frac{\left(\frac{h}{l}-\frac{1}{I_2}\right)\left(\frac{1}{I_1}-\frac{1}{I_3}\right)}{\left(\frac{h}{l}-\frac{1}{I_3}\right)\left(\frac{1}{I_1}-\frac{1}{I_2}\right)}\sinh^2\varphi\right\}}, \pm\sqrt{-1}\sqrt{\frac{2l\left(\frac{1}{I_2}-\frac{h}{l}\right)}{\frac{1}{I_2}-\frac{1}{I_3}}}\cosh\varphi\right), \notag
\end{align}
where $\varphi$ moves from $-\infty$ to $+\infty$. It is easy to see that ${\tau_{\pm}^{\prime}}^{(1)}$ are included in the same regular fibre of $\pi_F: F\rightarrow P_3(\mathbb{C})$ near the singular locus $a=d$. As to the behavior of ${\tau_{\pm}^{\prime}}^{(1)}$ where $\varphi$ approaches to the infinity, we have $(x:y:z:w)\rightarrow (-\sqrt{b-c}:\pm\sqrt{c-a}:\sqrt{a-b}:0)$ as $\varphi\rightarrow +\infty$ and $(x:y:z:w)\rightarrow (\sqrt{b-c}:\pm\sqrt{c-a}:\sqrt{a-b}:0)$ as $\varphi\rightarrow -\infty$. The arcs $\tau_{\pm}^{(1)}$, ${\tau_{\pm}^{\prime}}^{(1)}$ are connected at these infinity points as in Figure 1. 

\medskip

\begin{picture}(400,100)
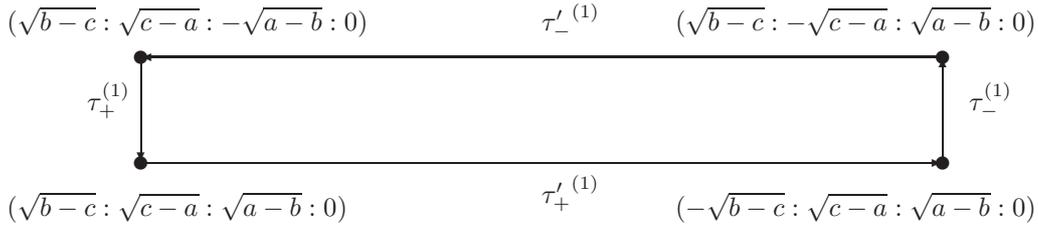
\put(120,90){Figure 1. The arcs $\tau_{\pm}^{(1)}$ and ${\tau_{\pm}^{\prime}}^{(1)}$}
\put(0,0){$(\sqrt{b-c}:\sqrt{c-a}:\sqrt{a-b}:0)$}
\put(250,0){$(-\sqrt{b-c}:\sqrt{c-a}:\sqrt{a-b}:0)$}
\put(250,70){$(\sqrt{b-c}:-\sqrt{c-a}:\sqrt{a-b}:0)$}
\put(0,70){$(\sqrt{b-c}:\sqrt{c-a}:-\sqrt{a-b}:0)$}
\put(50,20){\circle*{5}}
\put(350,20){\circle*{5}}
\put(350,60){\circle*{5}}
\put(50,60){\circle*{5}}
\put(50,20){\vector(1,0){300}}
\put(350,20){\vector(0,1){40}}
\put(350,60){\vector(-1,0){300}}
\put(50,60){\vector(0,-1){40}}
\put(200,5){${\tau_+^{\prime}}^{(1)}$}
\put(360,40){$\tau_-^{(1)}$}
\put(200,70){${\tau_-^{\prime}}^{(1)}$}
\put(30,40){$\tau_+^{(1)}$}
\end{picture}

\medskip

\noindent Note that there is no other intersection among $\tau_{\pm}^{(1)}$ and ${\tau_{\pm}^{\prime}}^{(1)}$ than these four points and that ${\tau_{\pm}^{\prime}}^{(1)}$ do not meet $\sigma_1^{(1)}$ and $\sigma_2^{(1)}$. We denote the multiplication $\tau_+^{(1)}\cdot{\tau_+^{\prime}}^{(1)}\cdot\tau_-^{(1)}\cdot{\tau_-^{\prime}}^{(1)}$ by $\tau^{(1)}$. In particular, $\tau^{(1)}$ is a closed arc in the regular fibre of $\pi_F:F \rightarrow P_3(\mathbb{C})$. 

In the first homology group $H_1(\pi_F^{-1}(p_0),\mathbb{Z})=G^{\prime}_{p_0}$, the cycles $\sigma_1^{(1)}$ and $\sigma_2^{(1)}$ are in the same homology class, which form a basis of this homology group together with $\tau^{(1)}$. Thus, calculating the period integrals ${\displaystyle \int_{\tau^{(1)}}\eta}$ and ${\displaystyle \int_{\sigma_j^{(1)}}\eta}$, $j=1$ or $j=2$, we can obtain a basis of $H^1(\pi_F^{-1}(p_0),\mathbb{Z})={G_{p_0}^{\prime}}^{\ast}$. 

We can express the period integral ${\displaystyle -\frac{1}{2\pi}\int_{\tau^{(1)}}\eta}$ by using an elliptic integral: 
\begin{theorem}
The period integral ${\displaystyle -\frac{1}{2\pi}\int_{\tau^{(1)}}\eta}$ of the one-form $\eta$ along the cycle $\tau^{(1)}$ is calculated as 
\begin{equation}
-\frac{1}{2\pi}\int_{\tau^{(1)}}\eta=S(c,b,a,d)=-\frac{1}{3\pi}\sqrt{\frac{2}{l}}\frac{1}{\sqrt{(d-a)(c-b)}}\mathcal{K}\left(\frac{(d-c)(b-a)}{(d-a)(b-c)}\right). \notag 
\end{equation}
\end{theorem}

\noindent\textbf{Proof.}\; 
We calculate the integrals of the one-form ${\displaystyle \eta=\frac{\mathsf{d}p_2}{3\left(\frac{1}{I_3}-\frac{1}{I_1}\right)p_3p_1}}$ along the four real arcs $\tau_{\pm}^{(1)}$ and ${\tau_{\pm}^{\prime}}^{(1)}$. For $\tau_{\pm}^{(1)}$, we have 
\begin{align}
\int_{\tau_+^{(1)}}\eta = &2\int_{\sqrt{\frac{2l\left(\frac{1}{I_1}-\frac{h}{l}\right)}{\frac{1}{I_1}-\frac{1}{I_2}}}}^{+\infty} \frac{\mathsf{d}p_2}{3\left(\frac{1}{I_3}-\frac{1}{I_1}\right)p_3p_1} \notag \\
= &\frac{1}{3}\sqrt{\frac{2}{l}}\frac{1}{\sqrt{\left(\frac{1}{I_3}-\frac{1}{I_1}\right)\left(\frac{h}{l}-\frac{1}{I_2}\right)}}\int_0^{+\infty}\frac{\mathsf{d}\varphi}{\sqrt{1+\frac{\left(\frac{h}{l}-\frac{1}{I_1}\right)\left(\frac{1}{I_3}-\frac{1}{I_2}\right)}{\left(\frac{h}{l}-\frac{1}{I_2}\right)\left(\frac{1}{I_3}-\frac{1}{I_1}\right)}\sinh^2\varphi}}. \notag
\end{align}
Putting $\sqrt{-1}x=\sinh\varphi$, we have 
\[
\int_{\tau_+^{(1)}}\eta=\frac{1}{3}\sqrt{\frac{2}{l}}\frac{1}{\sqrt{\left(\frac{h}{l}-\frac{1}{I_2}\right)\left(\frac{1}{I_1}-\frac{1}{I_3}\right)}}\int_0^{+\infty\sqrt{-1}}\frac{\mathsf{d}x}{\sqrt{(1-x^2)\left(1-\frac{\left(\frac{h}{l}-\frac{1}{I_1}\right)\left(\frac{1}{I_3}-\frac{1}{I_2}\right)}{\left(\frac{h}{l}-\frac{1}{I_2}\right)\left(\frac{1}{I_3}-\frac{1}{I_1}\right)}x^2\right)}}. 
\]
Note that the integral path is taken on the imaginary axis. Further, we have 
\begin{align}
\int_{\tau_+^{(1)}}\eta &=\frac{1}{3}\sqrt{2}{l}\frac{1}{\sqrt{(d-b)(a-c)}}\int_0^{+\infty\sqrt{-1}} \frac{\mathsf{d}x}{\sqrt{(1-x^2)\left(1-\frac{(d-a)(c-b)}{(d-b)(c-a)}x^2\right)}} \notag \\
&= \frac{1}{3}\sqrt{2}{l}\frac{1}{\sqrt{(d-b)(a-c)}}\left\{\int_0^{\sqrt{\frac{(d-b)(c-a)}{(d-a)(c-b)}}}\frac{\mathsf{d}x}{\sqrt{(1-x^2)\left(1-\frac{(d-a)(c-b)}{(d-b)(c-a)}x^2\right)}}\right. \notag \\
&\qquad\qquad\qquad \left. +\int_{\sqrt{\frac{(d-b)(c-a)}{(d-a)(c-b)}}}^{+\infty\sqrt{-1}}\frac{\mathsf{d}x}{\sqrt{(1-x^2)\left(1-\frac{(d-a)(c-b)}{(d-b)(c-a)}x^2\right)}}\right\} \notag \\
&= \frac{1}{3}\sqrt{\frac{2}{l}} \left\{ \frac{1}{\sqrt{(d-a)(b-c)}}\int_0^1\frac{\mathsf{d}x}{\sqrt{(1-x^2)\left(1-\frac{(d-b)(c-a)}{(d-a)(c-b)}x^2\right)}}\right. \notag \\
&\qquad\qquad\qquad \left. -\frac{1}{\sqrt{(d-b)(a-c)}} \int_0^1\frac{\mathsf{d}x}{\sqrt{(1-x^2)\left(1-\frac{(d-a)(c-b)}{(d-b)(c-a)}x^2\right)}}\right\} \notag \\
&=\frac{1}{3}\sqrt{\frac{2}{l}}\left\{\frac{1}{\sqrt{(d-a)(b-c)}}\mathcal{K}\left(\frac{(d-b)(c-a)}{(d-a)(c-b)}\right)+\frac{1}{\sqrt{(d-b)(a-c)}}\mathcal{K}\left(\frac{(d-a)(c-b)}{(d-b)(c-a)}\right)\right\}. \notag 
\end{align}
Thus, by \eqref{s_3-formulae}, we obtain 
\begin{align}
-\frac{1}{2\pi}\int_{{\tau_+}^{(1)}}\eta&=\frac{1}{2}\left(S(b,c,a,d)+S(a,c,b,d)\right) \notag \\
&=\frac{1}{2}\left(S(b,a,c,d)+S(a,b,c,d)\right). \notag 
\end{align}
Clearly, we have ${\displaystyle\int_{\tau_+^{(1)}}\eta=\int_{\tau_-^{(1)}}\eta}$ and ${\displaystyle \int_{{\tau_+^{\prime}}^{(1)}}\eta=-\int_{{\tau_-^{\prime}}^{(1)}}\eta}$ from the choice of the arcs. Thus, the integral ${\displaystyle {-\frac{1}{2\pi}}\int_{\tau^{(1)}}\eta}$ is calculated as 
\[
-\frac{1}{2\pi}\int_{\tau^{(1)}}\eta=-\frac{1}{\pi}\int_{\tau_+^{(1)}}\eta=S(b,a,c,d)+S(a,b,c,d). 
\]

To complete the proof, we show the following important formula. 
\begin{proposition}\label{connection-formula}
The following formula holds for the function $S$: 
\[
S(a,b,c,d)+S(b,a,c,d)=S(c,b,a,d). 
\]
\end{proposition}

For simplicity, we write as $S_1=S(a,b,c,d)$, $S_2=S(b,a,c,d)$, $S_3=S(c,b,a,d)$. The proof of the proposition can be performed by using the connection formula of a special Gau{\ss} hypergeometric equation. To describe this, we mention the relation between the complete elliptic integral ${\displaystyle \mathcal{K}(\lambda)=\int_0^1\frac{\mathsf{d}x}{\sqrt{(1-x^2)(1-\lambda x^2)}}}$ of the first kind and a Gau{\ss} hypergeometric equation. In fact, by means of Euler integral representation of the Gau{\ss} hypergeometric function (see e.g. \cite[Ch. 12. \S 12.5, p.494]{zoladek}, $\mathcal{K}(\lambda)$ can be written as 
\[
\mathcal{K}(\lambda)=\frac{\pi}{2}F\left(\frac{1}{2},\frac{1}{2},1;\lambda\right)=\frac{n}{2}\sum_{n=0}^{\infty}\left(\frac{(2n)!}{2^{2n}(n!)^2}\right)^2\lambda^n. 
\]
This shows that $\mathcal{K}(\lambda)$ satisfies the Gau{\ss} hypergeometric equation 
\begin{align}\label{gauss}
(1-\lambda)\lambda \frac{\mathsf{d}f}{\mathsf{d}\lambda^2}+(1-2\lambda)\frac{\mathsf{d}f}{\mathsf{d}\lambda}-\frac{1}{4}f=0. 
\end{align}
According to the description of the connection formula for this equation by \cite[7.405-7.406, pp.167-169]{caratheodory}, we consider the function ${\displaystyle F(\lambda):=F\left(\frac{1}{2},\frac{1}{2},1;\lambda\right)}$ and 
\[
F^{\ast}(\lambda)=F^{\ast}\left(\frac{1}{2},\frac{1}{2},1;\lambda\right)=4\sum_{n=1}^{\infty}\left(\frac{(2n)!}{2^{2n}(n!)^2}\right)^2\left(1-\frac{1}{2}+\frac{1}{3}-\cdots+\frac{1}{2n-1}-\frac{1}{2n}\right)\lambda^n. 
\]
Note that $F$ and $F^{\ast}$ are holomorphic around $\lambda=0$. Then, we set 
\begin{align}\label{gauss-functions}
\varphi_1&=F(\lambda)=\frac{1}{\sqrt{1-\lambda}}F\left(\frac{\lambda}{\lambda-1}\right), \notag \\
\varphi_3&=F(1-\lambda)=\frac{1}{\sqrt{\lambda}}F\left(\frac{\lambda-1}{\lambda}\right), \notag \\
\varphi_5&=\frac{1}{\sqrt{-\lambda}}F\left(\frac{1}{\lambda}\right)=\frac{1}{\sqrt{1-\lambda}}F\left(\frac{1}{\lambda-1}\right), \notag \\
\varphi_2^{\ast}&=F(\lambda)\log\lambda+F^{\ast}(\lambda) \notag \\
&=\frac{1}{\sqrt{1-\lambda}}\left\{F\left(\frac{\lambda}{\lambda-1}\right)\log\left(\frac{\lambda}{\lambda-1}\right)+F^{\ast}\left(\frac{\lambda}{\lambda-1}\right)\right\}, \notag \\
\varphi_4^{\ast}&=F(1-\lambda)\log(1-\lambda)+F^{\ast}(1-\lambda)\notag \\
&=\frac{1}{\sqrt{\lambda}}\left\{F\left(\frac{\lambda-1}{\lambda}\right)\log\left(\frac{1-\lambda}{\lambda}\right)+F^{\ast}\left(\frac{\lambda-1}{\lambda}\right)\right\}, \notag \\
\varphi_6^{\ast}&=\frac{1}{\sqrt{-\lambda}}\left\{F\left(\frac{1}{\lambda}\right)\log(-\lambda)-F^{\ast}\left(\frac{1}{\lambda}\right)\right\} \notag \\
&=\frac{1}{\sqrt{1-\lambda}}\left\{F\left(\frac{1}{1-\lambda}\right)\log(1-\lambda)-F^{\ast}\left(\frac{1}{1-\lambda}\right)\right\}.  
\end{align}
The pairs $\left(\varphi_1,\varphi_2^{\ast}\right)$, $\left(\varphi_3,\varphi_4^{\ast}\right)$, and $\left(\varphi_5,\varphi_6^{\ast}\right)$ give bases of the solution space of the Gau{\ss} hypergeometric equation \eqref{gauss} around $\lambda=0$, $\lambda=1$, and $\lambda=\infty$, respectively. Between two of them, we have the following connection formula: 
\begin{align}
\begin{cases}
\varphi_1=\frac{1}{\pi}\left(\varphi_3\log 16-\varphi_4^{\ast}\right), \\
\varphi_2^{\ast}=\frac{1}{\pi}\left\{\left((\log16)^2-\pi^2\right)\varphi_3-\varphi_4^{\ast}\log16\right\}=\varphi_1\log 16-\pi\varphi_3, 
\end{cases}\notag
\end{align}
\begin{align}
\begin{cases}
\varphi_1=\frac{1}{\pi}\left(\varphi_5\log 16+\varphi_6^{\ast}\right), \\
\varphi_2^{\ast}=\frac{1}{\pi}\left\{\left(\left(\log 16+\pi\sqrt{-1}\right)\log 16 -\pi^2\right)\varphi_5+\varphi_6^{\ast}\left(\log 16 +\pi\sqrt{-1}\right)\right\}=\varphi_1\left(\log 16+\pi\sqrt{-1}\right)-\pi\varphi_5.
\end{cases}\notag
\end{align}
See \cite[7.405-7.406, pp.167-169]{caratheodory}. From this connection formula, we see that $\varphi_1\sqrt{-1}+\varphi_3=\varphi_5$, which shows $S_1+S_2=S_3$. Note that 
\begin{align}
S_1&=-\frac{1}{3\pi}\sqrt{\frac{2}{l}}\frac{1}{\sqrt{(d-c)(a-b)}}\mathcal{K}\left(\frac{(d-a)(b-c)}{(d-c)(b-a)}\right)=-\frac{1}{3\pi}\sqrt{\frac{2}{l}}\frac{1}{\sqrt{(d-c)(a-b)}}\varphi_1(\lambda), \notag \\
S_2&=-\frac{1}{3\pi}\sqrt{\frac{2}{l}}\frac{1}{\sqrt{(d-c)(b-a)}}\mathcal{K}\left(\frac{(d-b)(a-c)}{(d-c)(a-b)}\right)=-\frac{1}{3\pi}\sqrt{\frac{2}{l}}\frac{1}{\sqrt{(d-c)(b-a)}}\varphi_3(\lambda), \notag \\
S_3&=-\frac{1}{3\pi}\sqrt{\frac{2}{l}}\frac{1}{\sqrt{(d-a)(c-b)}}\mathcal{K}\left(\frac{(d-c)(b-a)}{(d-a)(b-c)}\right)=-\frac{1}{3\pi}\sqrt{\frac{2}{l}}\frac{1}{\sqrt{(d-c)(b-a)}}\varphi_5(\lambda), \notag 
\end{align}
where ${\displaystyle \lambda=\frac{(d-a)(b-c)}{(d-c)(b-a)}}$. 
$\hfill \blacksquare$

By means of the previous proposition, the proof of the theorem is ended. 
$\hfill \blacksquare$

\section{Monodromy and global behavior of the Birkhoff normal forms}
In this section, we consider the global behavior of the Birkhoff normal forms, which can be found in relation to the monodromy of the naive elliptic fibration. We calculate the monodromy of the elliptic fibration $\pi_F:F\rightarrow P_3(\mathbb{C})$, using that of Gau{\ss} hypergeometric equation \eqref{gauss}. The two period integrals, ${\displaystyle -\frac{1}{2\pi}\int_{\sigma^{(1)}}\eta=S(a,b,c,d)}$ and ${\displaystyle -\frac{1}{2\pi}\int_{\tau^{(1)}}\eta=S(c,b,a,d)}$ of the one-form $\eta$ along the basis $\sigma^{(1)}$ and $\tau^{(1)}$ of the first homology group $H_1\left(\pi_F^{-1}(p_0),\mathbb{Z}\right)$ of a regular fibre $\pi_F^{-1}(p_0)$, where $p_0$ is near to the singular locus $a=d$, form a basis of the first cohomology group $H^1\left(\pi_F^{-1}(p_0),\mathbb{C}\right)$. We take a basis of the cohomology group $H^1\left(\pi_F^{-1}(p_0),\mathbb{C}\right)$ for a regular fibre $\pi_F^{-1}(p_0)$ over $p_0$ which is near to each component of the singular loci $D:\{a=b\}+ \{a=c\}+ \{a=d\}+ \{b=c\}+ \{b=d\}+ \{c=d\}$. To do this, we use the symmetry of the naive elliptic fibration and that of the function $S(a,b,c,d)$ with respect to the symmetric group $\mathfrak{S}_4$ acting on the base space $P_3(\mathbb{C}): (a:b:c:d)$ as the permutations of the four letters $a,b, c, d$.  In fact, we have the following list of equalities among $S(\sigma(a,b,c,d))$, for $\sigma\in\mathfrak{S}_4$, which can be obtained by using the equalities \eqref{gauss-functions} of the Gau{\ss} hypergeometric function $F$: 
\begin{align}\label{covariance}
\begin{array}{rllll}
S_1&=S(a,b,c,d)&=S(a,c,b,d)&=S(b,a,d,c)&=S(b,d,a,c)  \\
&=S(c,a,d,b)&=S(c,d,a,b)&=S(d,b,c,a)&=S(d,c,b,a),  \\
S_2&=S(b,a,c,d)&=S(b,c,a,d)&=S(a,b,d,c)&=S(a,d,b,c)  \\
&=S(c,b,d,a)&=S(c,d,b,a)&=S(d,a,c,b)&=S(d,c,a,b),  \\
S_3&=S(c,b,a,d)&=S(c,a,b,d)&=S(b,c,d,a)&=S(b,d,c,a) \\
&=S(a,c,d,b)&=S(a,d,c,b)&=S(d,a,b,c)&=S(d,b,a,c). 
\end{array}
\end{align}
This means that the function $S(a,b,c,d)$ is invariant with respect to the dihedral group generated by $(bc)$ and $(abdc)$, which is isomorphic to $\mathbb{Z}_2\ltimes \mathbb{Z}_4$. 

The naive elliptic fibration $\pi_F:F\rightarrow P_3(\mathbb{C})$ is invariant with respect to $\mathfrak{S}_4$ acting on $P_3(\mathbb{C})\times P_3(\mathbb{C}): ((x:y:z:w), (a:b:c:d))$ as \[
\sigma\left((x:y:z:w), (a:b:c:d)\right)=\left((\sigma x:\sigma y:\sigma z:\sigma w), (\sigma a:\sigma b:\sigma c:\sigma d)\right),
\]
for $\sigma \in\mathfrak{S}_4$. The irreducible components $b=c$, $b=d$, $a=c$, $c=d$, $a=b$ of the singular loci $D$ can be obtained from $a=d$ for instance by the action of $(ab)\cdot(cd)$, $(abc)$, $(acd)$, $(acb)$, $(abd)$, respectively. The basis $S_3=S(c,b,a,d)$, $S_1=S(a,b,c,d)$ of the cohomology group $H^1(\pi_F^{-1}(p_0),\mathbb{C})$ for $p_0$ near to $a=d$ is mapped by the action of $(ab)\cdot(cd)$, $(abc)$, $(acd)$, $(acb)$, $(abd)$ to the bases $S_3=S(d,a,b,c)$, $S_1=S(b,a,d,c)$; $S_1=S(a,c,b,d)$, $S_2=S(b,c,a,d)$; $S_1=S(d,b,c,a)$, $S_2=S(c,b,d,a)$; $S_2=S(b,a,c,d)$, $S_3=S(c,a,b,d)$; $S_2=S(c,d,b,a)$, $S_3=S(b,d,c,a)$ of the first cohomology group of the regular fibres near the components $b=c$, $b=d$, $a=c$, $c=d$, $a=b$, respectively. 

\medskip

We calculate the local monodromy of the fibration $\pi_F:F\rightarrow P_3(\mathbb{C})$ around each component of the singular loci $D$ with respect to the above basis. 
\begin{itemize}
\item Around the component $\{a=d\}+\{b=c\}$ of $D$, we have the basis $S_3, S_1$ of the first cohomology group of the regular fibre near this component. We take a real closed arc $\alpha_1$ in $P_3(\mathbb{C})\setminus \mathrm{Supp}(D)$, which is homotopic to the arc $a=d_0+\epsilon e^{\sqrt{-1}\theta},\; b=b_0,\; c=c_0,\; d=d_0$ or the one $a=a_0,\; b=c_0+\epsilon e^{\sqrt{-1}\theta},\; c=c_0,\; d=d_0$. Then, the lambda-function ${\displaystyle \lambda=\frac{(d-a)(b-c)}{(d-c)(b-a)}}$ can be seen to move as $\lambda=\epsilon e^{\sqrt{-1}\theta}\lambda_0$, ${\displaystyle \lambda_0=\frac{(d_0-a_0)(b_0-c_0)}{(d_0-c_0)(b_0-a_0)}}$, where $\theta: 0\rightarrow 2\pi$. Here, $a_0$, $b_0$, $c_0$, $d_0$ are suitable fixed complex numbers. The formulae \eqref{gauss-functions} implies that the basis $\varphi_1$, $\varphi_2^{\ast}$ of the solutions of the Gau{\ss} hypergeometric equation \eqref{gauss} is analytically continuated along the above arc as 
\begin{align}
\begin{bmatrix}
\varphi_1(\lambda) \\
\varphi_2^{\ast}(\lambda)
\end{bmatrix}\mapsto 
\begin{bmatrix}
\varphi_1(\lambda) \\
2\pi\sqrt{-1}\varphi_1(\lambda)+\varphi_2^{\ast}(\lambda)
\end{bmatrix}. \notag 
\end{align}
Using the connection formula $\varphi_2^{\ast}=\varphi_1(\log 16+\pi\sqrt{-1})-\pi \varphi_5$, we can see that the basis $S_3, S_1$ are analytically continuated along $\alpha_1$ as 
\[
\begin{bmatrix}
S_3 \\ S_1
\end{bmatrix}\mapsto 
\begin{bmatrix}
1 & 2 \\ 0 & 1 
\end{bmatrix}
\begin{bmatrix}
S_3 \\ S_1
\end{bmatrix}.
\]
\item Around the component $\{b=d\}+\{a=c\}$ of $D$, we take the basis $S_1$, $S_2$ of the first cohomology group of the regular fibre near the component. Consider a real closed arc $\alpha_2$ in $P_3(\mathbb{C})\setminus \mathrm{Supp}(D)$, which is homotopic to the arc $a=a_0,\; b=d_0,+\epsilon e^{\sqrt{-1}\theta}\; c=c_0,\; d=d_0$ or the one $a=c_0+\epsilon e^{\sqrt{-1}\theta},\; b=b_0,\; c=c_0,\; d=d_0$. Then, ${\displaystyle 1-\lambda=\frac{(d-b)(a-c)}{(d-c)(a-b)}}$ can be assumed to move as $1-\lambda=\epsilon e^{\sqrt{-1}\theta}$, ${\displaystyle \lambda_0=\frac{(d_0-b_0)(a_0-c_0)}{(d_0-c_0)(a_0-b_0)}}$, where $\theta:0\rightarrow 2\pi$. From \eqref{gauss-functions}, the basis $\varphi_3$, $\varphi_4^{\ast}$ of the solution space of the Gau{\ss} hypergeometric equation is analytically continuated along the above arc as 
\begin{align}
\begin{bmatrix}
\varphi_3(\lambda) \\
\varphi_4^{\ast}(\lambda)
\end{bmatrix}\mapsto 
\begin{bmatrix}
\varphi_3(\lambda) \\
2\pi\sqrt{-1}\varphi_3(\lambda)+\varphi_4^{\ast}(\lambda)
\end{bmatrix}. \notag 
\end{align}
From the connection formula ${\displaystyle \varphi_1=\frac{1}{\pi}\left(\varphi_3\log 16 -\varphi_4^{\ast}\right)}$, we see that the basis $S_1$, $S_2$ are analytically continuated along $\alpha_2$ as 
\[
\begin{bmatrix}
S_1 \\ S_2
\end{bmatrix}\mapsto 
\begin{bmatrix}
1 & -2 \\ 0 & 1 
\end{bmatrix}
\begin{bmatrix}
S_1 \\ S_2
\end{bmatrix}.
\]
\item Around the component $\{c=d\}+\{a=b\}$ of $D$, we have the basis $S_2$, $S_3$ of the first cohomology group of the regular fibre near the component. Take a real closed arc $\alpha_3$ in $P_3(\mathbb{C})\setminus \mathrm{Supp}(D)$, which is homotopic to the arc $a=a_0$, $b=b_0$, $c=d_0+\epsilon e^{\sqrt{-1}\theta}$, $d=d_0$; or the one $d=d_0$; and $a=b_0+\epsilon e^{\sqrt{-1}\theta}$, $b=b_0$, $c=c_0$, $d=d_0$. Then, ${\displaystyle \frac{1}{\lambda}=\frac{(d-c)(b-a)}{(d-a)(b-c)}}$ can be assumed to move as ${\displaystyle \frac{1}{\lambda}=\epsilon e^{\sqrt{-1}\theta}\frac{1}{\lambda_0}}$, ${\displaystyle \lambda_0=\frac{(d_0-a_0)(b_0-c_0)}{(d_0-c_0)(b_0-a_0)}}$, where $\theta: 0\rightarrow 2\pi$. From the basis $\varphi_5$, $\varphi_6^{\ast}$ of the solution space of the Gau{\ss} hypergeometric equation is analytically continuated along the above arc as 
\begin{align}
\begin{bmatrix}
\varphi_5(\lambda) \\
\varphi_6^{\ast}(\lambda)
\end{bmatrix}\mapsto 
\begin{bmatrix}
-\varphi_5(\lambda) \\
-2\pi\sqrt{-1}\varphi_5(\lambda)-\varphi_6^{\ast}(\lambda)
\end{bmatrix}. \notag 
\end{align}
From the connecting formulae, we have $\varphi_6^{\ast}=\pi\sqrt{-1}\varphi_3+\left(-\pi\sqrt{-1}+\log 16\right)\varphi_5$. Taking into account the fact that the analytic continuation of $\sqrt{(d-c)(b-a)}$ along the above arc $\alpha_3$ is given by the multiplication of $-1$, we find the analytic continuation of $S_2, S_3$ as 
\[
\begin{bmatrix}
S_2 \\ S_3
\end{bmatrix}\mapsto 
\begin{bmatrix}
1 & 2 \\ 0 & 1 
\end{bmatrix}
\begin{bmatrix}
S_2 \\ S_3
\end{bmatrix}.
\]
\end{itemize}

\medskip

Before dealing with the global monodromy, we describe the fundamental group of the complement $P_3(\mathbb{C})\setminus \mathrm{Supp}(D)$ of the singular locus $D$ of the fibration $\pi_F:F\rightarrow P_3(\mathbb{C})$. Note that the equations $a=b$, $a=c$, $a=d$, $b=c$, $b=d$, $c=d$ of the singular locus $D$ form the arrangement of affine hyperplanes in $V=\mathbb{C}^4: (a,b,c,d)$ of type $A_3$. As to the complement of the affine hyperplane arrangement, it is known that its fundamental group is the colored braid group \cite{brieskorn_2}. Applying this general argument to our setting, we can calculate the fundamental group as follows: \\
On $V$, there is the action of the Weyl group $\mathcal{W}=\mathfrak{S}_4$ whose elements permute the coordinates $a$, $b$, $c$, $d$. Obviously, its fixed-point-set is the union of the six hyperplanes $a=b$, $a=c$, $a=d$, $b=c$, $b=d$, $c=d$. Set 
\[
Y=V\setminus \left(\{a=b\}\cup\{a=c\}\cup\{a=d\}\cup\{b=c\}\cup\{b=d\}\cup\{c=d\}\right), 
\]
and $X:=Y/\mathcal{W}$. The quotient mapping $Y\rightarrow X$ is an unramified covering and we have the exact sequence 
\[
1\rightarrow \pi_1(Y)\rightarrow \pi_1(X)\rightarrow \mathcal{W}\rightarrow 1. 
\]
Further, the fundamental group $\pi_1(X)$ of the quotient space is the braid group generated by the generators $g_1$, $g_2$, $g_3$ with the relations $g_1g_2g_1=g_2g_1g_2$, $g_2g_3g_2=g_3g_2g_3$, $g_1g_3=g_3g_1$ (cf. \cite{brieskorn_1,brieskorn_2}). In the sense of geometry of braids, $g_1$, $g_2$, $g_3$ describe the simplest strands between the first and the second strings, the second and the third strings, the third and the fourth strings, respectively, for the braids with four strings. The group homomorphism $\pi_1(X)\rightarrow \mathcal{W}=\mathfrak{S}_4$ is given by the natural correspondence 
\[
g_1\mapsto (12),\quad g_2\mapsto (23),\quad g_3\mapsto (34). 
\]
By the above short exact sequence, we can realize $\pi_1(Y)$ as the kernel of the this homomorphism. More precisely, putting $h_{12}=g_1^2$, $h_{23}=g_2^2$, $h_{34}=g_3^2$, $h_{13}=g_1g_2^2g_1^{-1}=g_2^{-1}g_1^2g_2$, $h_{14}=g_1g_2g_3^2g_2^{-1}g_1^{-1}=g_3^{-1}g_1g_2^2g_1^{-1}g_3=g_3^{-1}g_2^{-1}g_1^2g_2g_1$, $h_{24}=g_2g_3^2g_2^{-1}=g_3^{-1}g_2^2g_3$, we can describe $\pi_1(Y)$ as the group generated by $h_{12}$, $h_{23}$, $h_{34}$, $h_{13}$, $h_{14}$, $h_{24}$ with the relations 
\begin{align}\label{braids}
h_{12}h_{23}h_{13}&=h_{23}h_{13}h_{12}=h_{13}h_{12}h_{23}, \notag \\
h_{23}h_{34}h_{24}&=h_{34}h_{24}h_{23}=h_{24}h_{23}h_{34}, \notag \\
h_{12}h_{24}h_{14}&=h_{24}h_{14}h_{12}=h_{14}h_{12}h_{24}, \notag \\
h_{34}h_{14}h_{13}&=h_{14}h_{13}h_{34}=h_{13}h_{34}h_{14}, \\
h_{12}h_{34}&=h_{34}h_{12}, \notag\\
h_{13}h_{23}^{-1}h_{24}h_{23}&=h_{23}^{-1}h_{24}h_{23}h_{13}, \notag \\
h_{23}h_{14}&=h_{14}h_{23}. \notag 
\end{align}

\medskip

Now, we consider the complement of the projective (hyper)plane arrangement. By taking the quotient through the fixed-point-free action of the group $\mathbb{C}^{\ast}$, we have the complement $P_3(\mathbb{C})\setminus \mathrm{Supp}(D)$: 
\begin{align}
\begin{array}{rcl}
\mathbb{C}^{\ast} & \rightarrow & Y\\
 & & \downarrow \\
 & & \!\!\!\!\!\! P_3 (\mathbb{C})\setminus \mathrm{Supp}(D)
\end{array} \notag 
\end{align}

The homotopy exact sequence of fibre bundles (cf. \cite{steenrod}) yields 
\[
1\rightarrow \pi_1(\mathbb{C}^{\ast})\rightarrow \pi_1(Y)\rightarrow \pi_1(P_3(\mathbb{C})\setminus \mathrm{Supp}(D))\rightarrow 1,
\]
so that $\pi_1(P_3(\mathbb{C})\setminus \mathrm{Supp}(D))\cong\pi_1(Y)/\pi_1(\mathbb{C}^{\ast})$. In order to determine the fundamental group $\pi_1(P_3(\mathbb{C})\setminus \mathrm{Supp}(D))$, we have to find a generator of $\pi_1(\mathbb{C}^{\ast})\cong \mathbb{Z}$. Since all the hyperplanes $a=b$, $a=c$, $a=d$, $b=c$, $b=d$, $c=d$ in $V$ pass through the origin, the generators of $\pi_1(\mathbb{C}^{\ast})$ can be realized as a multiple of the six elements in $\pi_1(Y)$ presented by closed arcs which enclose the hyperplanes. Further, the generator is in the centre of the fundamental group $\pi_1(Y)$. This follows from the following theorem, by means of Zariski's Theorem (cf. \cite[Chapter 5, \S 5.3]{orlik-terao}). 
\begin{theorem}[Randell]
Let $l_1, \ldots, l_n$ be affine complex lines in $\mathbb{C}^2$ which are defined by linear equations with real coefficients and which pass through the origin. Figure 2 describes the real section of the line arrangement. Denote the generators of $\pi_1\left(\mathbb{C}^2\setminus \cup_{j=1}^n l_j\right)$ represented by closed arcs around $l_j$ by $\gamma_j$ and $\gamma_j^{\prime}$ as in Figure 2. Then, we have 
\[
\gamma_1\cdots\gamma_n=\gamma_2\cdots\gamma_n\gamma_1=\cdots=\gamma_n\gamma_1\cdots\gamma_{n-1}
\]
and 
\[
\gamma_j^{\prime}=\gamma_n^{-1}\cdots\gamma_{j-1}^{-1}\gamma_j\gamma_{j+1}\cdots\gamma_n. 
\]
\begin{center}
\begin{picture}(250,110)
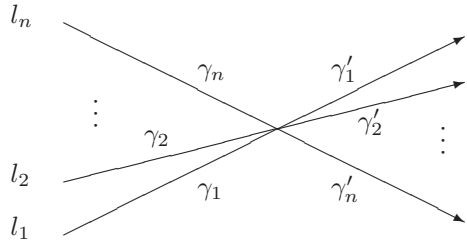
\put(0,100){Figure 2. Lines in $\mathbb{C}^2$ passing through the origin} 
\put(20,0){\vector(2,1){150}}
\put(20,20){\vector(4,1){150}}
\put(30,40){\rotatebox{90}{$\cdots$}}
\put(160,30){\rotatebox{90}{$\cdots$}}
\put(20,80){\vector(2,-1){150}}
\put(0,0){$l_1$}
\put(0,20){$l_2$}
\put(0,80){$l_n$}
\put(70,15){$\gamma_1$}
\put(50,35){$\gamma_2$}
\put(70,60){$\gamma_n$}
\put(120,15){$\gamma_n^{\prime}$}
\put(130,40){$\gamma_2^{\prime}$}
\put(120,60){$\gamma_1^{\prime}$}
\end{picture}
\end{center}
\end{theorem}
For the proof, see \cite{randell} or \cite{orlik-terao}, although \cite{orlik-terao} deals with the fundamental group of the complement of more general hyperplane arrangements. From this theorem, we see that $\gamma=\gamma_1\cdots\gamma_n$ commute with $\gamma_j$, since $\gamma\gamma_j=\gamma_j\gamma_{j+1}\cdots\gamma_n\gamma_1\cdots\gamma_{j-1}\gamma_j=\gamma_j\gamma$. The centre of $\pi_1(Y)$ is generated by $(g_1g_2g_3)^4$ (cf. \cite{artin}) and, further, we have the expression 
\[
(g_1g_2g_3)^4=h_{13}h_{12}h_{23}h_{34}h_{24}h_{14}. 
\]
Therefore, $\pi_1(\mathbb{C}^{\ast})$ is generated by $h_{13}h_{12}h_{23}h_{34}h_{24}h_{14}$. The fundamental group $\pi_1\left(P_3(\mathbb{C})\setminus \mathrm{Supp}(D)\right)$ can be realized as the group generated by $h_{12}, h_{13}, h_{14}, h_{23}, h_{24}, h_{34}$ with the relations \eqref{braids} and $h_{13}h_{12}h_{23}h_{34}h_{24}h_{14}=1$. 

\medskip

Here, we make  comments on the relation between the fundamental group of $P_3(\mathbb{C})\setminus \mathrm{Supp}(D)$ and that of the complement of the line arrangement on the (hyper)plane $E: a+b+c+d=0$, which is isomorphic to $P_2(\mathbb{C})$, given by  the same equations as $D$: $a=b$, $a=c$, $a=d$, $b=c$, $b=d$, $c=d$. Denote the induced divisor on the plane $E: a+b+c+d=0$ by $\overline{D}$. We have the following proposition. 
\begin{proposition}
We have the following isomorphism: 
\[
\pi_1(P_3(\mathbb{C})\setminus \mathrm{Supp}(D))\cong \pi_1(E\setminus \mathrm{Supp}(\overline{D})). 
\]
\end{proposition}

\noindent\textbf{Prrof.}\; We consider the blowing-up of $P_3(\mathbb{C}): (a:b:c:d)$ with the centre at $a=b=c=d$ as \cite{naruki-tarama}: $\Phi_B:B\rightarrow P_3(\mathbb{C})$. The hyperplane $E$ can be identified with the exceptional divisor through $\Phi_B$. Clearly, we have $P_3(\mathbb{C})\setminus \mathrm{Supp}(D)\cong B\setminus \left(E\cup\mathrm{Supp}(\widetilde{D})\right)$, where $\widetilde{D}$ is the proper transform of $D$ through $\Phi_B$. On the other hand, identifying all the points on a line in $P_3(\mathbb{C})$ which pass through the point $a=b=c=d$, we have the mapping $\tau_B:B\rightarrow E $, which is in fact a $P_1(\mathbb{C})$-fibre bundle. This structure of $P_1(\mathbb{C})$-fibre bundle is inherited to the complement $B\setminus \left(E\cup\mathrm{Supp}(\widetilde{D})\right)$, so that we have the $P_1(\mathbb{C})$-fibre bundle $B\setminus\left(E\cup\mathrm{Supp}(\widetilde{D})\right)\rightarrow E\setminus\mathrm{Supp}(\overline{D})$. By the homotopy exact sequence of fibre bundles \cite{steenrod}, we have the exact sequence $\pi_1(P_1(\mathbb{C}))\rightarrow \pi_1\left(B\setminus\left(E\cup \mathrm{Supp}\left(\widetilde{D}\right)\right)\right)\rightarrow \pi_1\left(E\setminus \mathrm{Supp}\left(\overline{D}\right)\right)\rightarrow 1$. The simple connectedness of $P_1(\mathbb{C})$ proves the proposition. 
$\hfill \blacksquare$

We draw the real section of the arrangement $\overline{D}$ on the plane $E\cong P_2(\mathbb{C})$ as in Figure 3. 
\begin{center}
\begin{picture}(300,250)
\thicklines
\put(25,50){\vector(1,0){250}}
\put(260,35){\vector(-2,3){130}}
\put(40,35){\vector(2,3){130}}
\put(150,25){\vector(0,1){210}}
\put(20,35){\vector(2,1){200}}
\put(80,135){\vector(2,-1){200}}
\put(0,55){$a=b$}
\put(245,25){$a=c$}
\put(135,15){$c=d$}
\put(225,135){$b=d$}
\put(175,235){$a=d$}
\put(60,140){$b=c$}
\put(100,0){Figure 3. $A_3$ configuration on $E$}
\put(153,27){$h_{34}$}
\put(110,40){$h_{12}$}
\put(90,80){$h_{24}$}
\put(60,90){$h_{14}$}
\put(120,120){$h_{23}$}
\put(225,90){$h_{13}$}
\end{picture}
\end{center}

In relation to the description of $\pi_1(P_3(\mathbb{C})\setminus \mathrm{Supp}(D))$ by the generators $h_{12}$, $h_{23}$, $h_{34}$, $h_{13}$, $h_{14}$, $h_{24}$, the corresponding closed arcs can be chosen as indicated in Figure 3. Note that the closed arc around a line is chosen as follows: \\
Let $l$ be the line in $\mathbb{C}^2: (u,v)$ given by $v=\alpha u$, where $\alpha$ is a real positive constant. We fix the orientation of $l$ as indicated in Figure 4, which describes the real section. Corresponding to this line with the orientation, we assign the closed arc 
\[
\begin{cases}
u=\epsilon e^{\sqrt{-1}\theta}, \\
v=0, 
\end{cases}
\]
where $\theta: 0\rightarrow 2\pi$ and $\epsilon$ is a sufficiently small real constant. 
\begin{center}
\begin{picture}(200,100)
\put(10,40){\vector(1,0){150}}
\put(50,10){\vector(0,1){75}}
\put(165,37){$\mathrm{Re}(u)$}
\put(45,90){$\mathrm{Im}(u)$}
\put(20,10){\vector(1,1){70}}
\put(20,25){\vector(2,1){100}}
\put(120,80){$l$}
\put(80,90){$\mathrm{Re}(v)$}
\put(40,45){$0$}
\put(50,40){\circle{30}}
\put(66,41){\vector(0,1){1}}
\put(34,39){\vector(0,-1){1}}
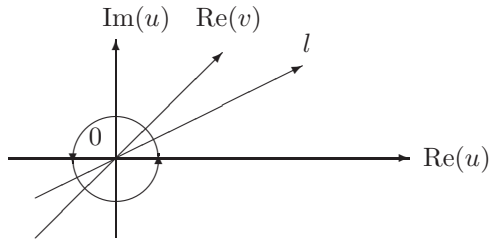
\put(60,0){Figure 4. Closed arc around $l$}
\end{picture}
\end{center}

\medskip

By the results of local monodromy and the connection formula $S_1+S_2=S_3$ in Proposition \ref{connection-formula}, we can calculate the global monodromy. Note that 
\[
\begin{bmatrix}
S_1 \\ S_2
\end{bmatrix}
=
\begin{bmatrix}
0 & 1 \\ 1 & -1
\end{bmatrix}
\begin{bmatrix}
S_3 \\ S_1
\end{bmatrix} 
\]
and 
\[
\begin{bmatrix}
S_2 \\ S_3
\end{bmatrix}
=
\begin{bmatrix}
1 & -1 \\ 1 & 0
\end{bmatrix}
\begin{bmatrix}
S_3 \\ S_1
\end{bmatrix}. 
\]
\begin{theorem}\label{thm6.3}
The basis of the first cohomology group for the regular fibre of the fibration $\pi_F: F\rightarrow P_3(\mathbb{C})$ is generated by the analytic continuations of $S_1$ and of $S_3$, which are proportional to the derivative of the inverse Birkhoff normal forms around the $p_1$- and $p_3$-axes, respectively. The monodromy of the fibration $\pi_F$ with respect to $S_1$ and $S_3$ is given by the correspondence of the generators $h_{12}$, $h_{13}$, $h_{14}$, $h_{23}$, $h_{24}$, $h_{34}$ of the fundamental group $\pi_1(P_3(\mathbb{C})\setminus \mathrm{Supp}(D))$ to the matrices in $SL(2,\mathbb{Z})$ as follows: 
\begin{align}\label{global-monodromy}
h_{12}, h_{34}\mapsto 
\begin{bmatrix}
1 & 2 \\ 0 & 1
\end{bmatrix}, 
h_{13}, h_{24}\mapsto 
\begin{bmatrix}
-1 & 2 \\ -2 & 3
\end{bmatrix}, 
h_{14}, h_{23}\mapsto 
\begin{bmatrix}
1 & 0 \\ -2 & 1
\end{bmatrix}. 
\end{align}
\end{theorem}

\begin{remark}
Here, we mention the relation of our analysis to the paper \cite{francoise-garrido-gallavotti_2}. 

According to \cite[\S VI]{francoise-garrido-gallavotti_2}, we consider the Birkhoff normal form around the $p_2$-axis under the condition $I_3 <I_1<I_2$. (Note that we have used another order $I_1<I_2<I_3$, but the result coincides if we make the permutation $(abc)$ or $(I_1I_2I_3)$.) We take the normalized Hamiltonian 
\[
Z=\frac{1}{4r^2}\frac{\frac{H}{l}-b}{c-b}=\frac{1}{4r^2}\frac{\frac{H}{l}-\frac{1}{I_2}}{\frac{1}{I_3}-\frac{1}{I_2}}, 
\]
where ${\displaystyle r^2=\frac{a-b}{c-b}}$, and considered it as a power series $Z=\frac{1}{4}F(x,r^2)$ with the parameter $r^2$. Denote the inverse as $x=Z\widehat{C}(Z)^2$, ${\displaystyle \widehat{C}(Z)^2=\sum_{n=0}^{\infty}\frac{P_n(r^2)}{n+1}Z^n}$, where $P_n$ is a polynomial in $r^2$ of degree $n$. One of the main result of \cite{francoise-garrido-gallavotti_2} is that the root of $P_n$ are on the unit circle in the Gau{\ss} plane. For the proof, the following symmetry property of $P_n$ is one of the essential conditions: 
\begin{align}\label{symmetry-coefficient}
r^{2n}P_n(\frac{1}{r^2})=P_n(r^2). 
\end{align}
In fact, this property of $P_n$ can be deduced from the covariance property \eqref{covariance} of the function $S$. Indeed, the derivative of the inverse Birkhoff normal form is given by 
\[
\frac{1}{4}\sum_{n=0}^{\infty}P_n(r^2)Z^n, 
\]
which is equal to $S(b,a,c,d)$. Therefore, we have, from $S(b,a,c,d)=S(b,c,a,d)$, 
\[
\sum_{n=0}^{\infty}P_n(r^2)Z^n=\sum_{n=0}^{\infty}P_n\left(\frac{1}{r^2}\right)(r^2Z)^n, 
\]
where we used the fact that $r^2$ is mapped to ${\displaystyle \frac{1}{r^2}}$ and $Z$ to $r^2Z$, by the action of the permutation $(ac)$. 
\end{remark}

\begin{remark}
In \cite{naruki-tarama}, several different elliptic fibrations are considered besides the naive elliptic fibration $\pi_F:F\rightarrow P_3(\mathbb{C})$ in relation to the free rigid body dynamics. As before, we consider the blowing-up $\Phi_B:B\rightarrow P_3(\mathbb{C})$ of $P_3(\mathbb{C}):(a:b:c:d)$ with the centre at $a=b=c=d$, and the projection $\tau_B:B\rightarrow E\subset P_3(\mathbb{C})$ to the exceptional set of $\Phi_B$. In fact, we have the following commutative diagram of elliptic fibrations: 
\begin{align}\label{fibrations}
\begin{array}{ccccccccccccccc}
F & \leftarrow & \Phi_B^{\ast}F & \cong & \tau_B^{\ast}\overline{F} & \rightarrow & \overline{F} & -\stackrel{4:1}{\cdots}\rightarrow  & T & \leftarrow & \tau_B^{\ast}T & \cong & \Phi_B^{\ast}W & \rightarrow & W \\
\downarrow & & \downarrow & & \downarrow & & \downarrow & & \downarrow & & \downarrow & & \downarrow & & \downarrow \\
P_3(\mathbb{C}) & \stackrel{\Phi_B}{\leftarrow} & B & = & B & \stackrel{\tau_B}{\rightarrow} & E & = & E & \stackrel{\tau_B}{\leftarrow} &B & = & B & \stackrel{\Phi_B}{\rightarrow} & P_3(\mathbb{C}) 
\end{array}
\end{align}

The arrows from above to down are elliptic fibrations: $\pi_F: F\rightarrow P_3(\mathbb{C})$, $\pi_{\Phi_B^{\ast}F}:\Phi_B^{\ast}F\rightarrow B$, $\pi_{\tau_B^{\ast}\overline{F}}: \tau_B^{\ast}\overline{F}\rightarrow B$, $\pi_{\overline{F}}:\overline{F}\rightarrow E$, $\pi_{T}:T\rightarrow E$, $\pi_{\tau_B^{\ast}T}:\tau_B^{\ast}T\rightarrow B$, $\pi_{\Phi_b^{\ast}W}: \Phi_B^{\ast}W\rightarrow B$, $\pi_W:W\rightarrow P_3(\mathbb{C})$. The horizontal arrows are bimeromorphisms except $\overline{T}-\cdots\rightarrow T$, which is a $4:1$ meromorphic mapping and $\tau_B:B\rightarrow E$, $\Phi_ B^{\ast}\overline{F}\rightarrow \overline{F}$, $\tau_B^{\ast}T\rightarrow T$, which are projections. The fibration $\pi_W: W\rightarrow P_3(\mathbb{C})$ and $\pi_T: T\rightarrow E$ are in Weierstra{\ss} normal form, while $\pi_{\overline{F}}:\overline{F}\rightarrow E$ is not in Weierstra{\ss} normal form. The fibration $\pi_{\overline{F}}: \overline{F}\rightarrow E$, which was not explicitly considered in \cite{naruki-tarama}, is naturally induced on $E$, since the fibration $\pi_F$ has the same fibre along the fibre of $\tau_B$. The singular loci of these elliptic fibrations are as follows: \\
\begin{itemize}
\item For $\pi_F:F\rightarrow P_3(\mathbb{C})$ and $\pi_W: W\rightarrow P_3(\mathbb{C})$, the singular fibre is the divisor $D=\{a=b\}+\{a=c\}+\{a=d\}+\{b=c\}+\{b=d\}+\{c=d\}$ on $P_3(\mathbb{C}): (a:b:c:d)$. 
\item For $\pi_{\Phi_B^{\ast}F}:\Phi_B^{\ast}F\rightarrow B$, $\pi_{\Phi_B^{\ast}W}:\Phi_B^{\ast}W\rightarrow B$, the total transform of $D$ through $\Phi_B:B\rightarrow P_3(\mathbb{C})$, i.e. $\widetilde{D}+E$, where $\widetilde{D}$ is the proper transform of $D$ through $\Phi_B$. 
\item For $\pi_{\tau_B^{\ast}\overline{F}}:\tau_B^{\ast}\overline{F}\rightarrow B$, $\pi_{\tau_B^{\ast}T}:\tau_B^{\ast}T\rightarrow B$, the proper transform $\widetilde{D}$. 
\item For $\pi_{\overline{F}}:\overline{F}\rightarrow E$, $\pi_T:T\rightarrow E$, the divisor $\overline{D}=\{a=b\}+\{a=c\}+\{a=d\}+\{b=c\}+\{b=d\}+\{c=d\}$ on $E=\{a+b+c+d=0\}\subset P_3(\mathbb{C}): (a:b:c:d)$. 
\end{itemize}

We set $P_3(\mathbb{C})^{\ast}=P_3(\mathbb{C})\setminus\{(1:1:1:1)\}$, $B^{\ast}=B\setminus E$, $B^{\prime}=B\setminus \mathrm{Supp}(\widetilde{D})$, $B^{\prime\prime}=B\setminus \mathrm{Supp}\left(\widetilde{D}+E\right)$, $P_3(\mathbb{C})^{\prime}=P_3(\mathbb{C})\setminus \mathrm{Supp}(D)$, $E^{\prime}=E\setminus \mathrm{Supp}(\widetilde{D})$. Then, we have $B^{\prime\prime}\subset B^{\prime}$, $B^{\prime}\setminus B^{\prime}\cong E$. Using the homotopy exact sequence of fibre bundles, we can show that these regular loci of $B^{\prime}$, $B^{\prime\prime}$, $E^{\prime}$ for the four elliptic fibrations $\pi_F$, $\pi_W$, $\pi_{\Phi_B^{\ast}F}$, $\pi_{\Phi_B^{\ast}W}$; for two fibrations  $\pi_{\tau_B^{\ast}\overline{F}}$, $\pi_{\tau_B^{\ast}T}$; and for two fibrations $\pi_{\overline{F}}$, $\pi_T$, respectively, have the isomorphic fundamental groups: 
\[
\pi_1(B^{\prime},\ast)\cong \pi_1(B^{\prime\prime},\ast)\cong \pi_1(E^{\prime},\ast). 
\]
Moreover, the $4:1$ meromorphic mapping $\overline{F}-\cdots\rightarrow T$ induces an isogeny on each regular fibres. By means of these facts, we can show that the monodromy representations of all the elliptic fibrations which appeared in \eqref{fibrations} are equivalent. 
\end{remark}

\begin{remark}
In \cite[Theorem 2]{naruki-tarama}, the bimeromorphism between $\pi_W:W\rightarrow P_3(\mathbb{C})$ and $\pi_{\widehat{W}}: \widehat{W}\rightarrow \widehat{B}$ should be understood as a biholomorphic mapping between the Zariski open set consisting of regular fibres of $\pi_W$ and that of $\pi_{\widehat{W}}$. For the Zariski open set of $\widehat{W}$, we need to subtract the fibres over the proper transforms of the exceptional set $E$ through $\Phi_B$. 
\end{remark}

\begin{remark}
In \cite{naruki}, the confluence of singular fibres in elliptic fibrations is discussed from the view point of monodromy. In particular, the monodromy matrices for the confluence of three singular fibres of type $\mathrm{I}_2$ to that of type $\mathrm{I}_0^{\ast}$ in Kodaira's notation are determined as 
\[
\begin{bmatrix}1 & 2 \\ 0 & 1 \end{bmatrix}, \quad \begin{bmatrix}1 & 0 \\ -2 & 1\end{bmatrix}, \quad \begin{bmatrix}-1 & 2 \\ -2 & 3\end{bmatrix}, 
\]
up to simultaneous conjugations by $SL(2,\mathbb{Z})$. This is checked by an easy computation as follows: 
\[
\begin{bmatrix}1 & 2 \\ 0 & 1 \end{bmatrix}\begin{bmatrix}1 & 0 \\ -2 & 1\end{bmatrix}\begin{bmatrix}-1 & 2 \\ -2 & 3\end{bmatrix}=\begin{bmatrix}-1 & 0 \\ 0 &-1\end{bmatrix}. 
\]
See \cite[\S 8, Table 5]{naruki}. 

On the other hand, the modification of the total and the base spaces of the fibration $\pi_T:T\rightarrow E$ and $\pi_{\Phi_B^{\ast}W}: \Phi_B^{\ast}W\rightarrow B$ are performed to obtain smooth and flat fibrations which admit only singular fibres in Kodaira's list of singular fibres for elliptic surfaces in \cite{naruki-tarama}. For the fibration $\pi_T: T\rightarrow E$, the blowing-up $\widehat{E}\rightarrow E$ with the separate centres at the four points $(a:b:c:d)=(-3:1:1:1), (1:-3:1:1), (1:1:-3:1), (1:1:1:-3)$, where three of the six lines intersect, is important to obtain such a desired fibration. See Figure 3. For the finally obtained elliptic fibration $\pi_{\widehat{T}}:\widehat{T}\rightarrow \widehat{E}$, the singular fibres on the exceptional sets through the blowing-up $\widehat{E}\rightarrow E$ are in general of type $\mathrm{I}_0^{\ast}$ in Kodaira's notation. Note that the monodromy matrix for the singular fibre of type $\mathrm{I}_0^{\ast}$ is 
\[
\begin{bmatrix}-1 & 0 \\ 0 &-1\end{bmatrix}. 
\]

This configuration of singular fibres is compatible with the result in Theorem \ref{thm6.3}. In fact, we have seen in Theorem \ref{thm6.3} that the global monodromy is given by the correspondence \eqref{global-monodromy}. A similar comparison can be performed also on the fibration $\pi_{\widehat{W}}:\widehat{W}\rightarrow \widehat{B}$. 
\end{remark}

\noindent\textbf{Acknowledgment:} This work was done during the stay of the second author at Laboratoire Jacques-Louis Lions, Universit\'{e} Pierre-Marie Curie, with the support by \textit{Research in Paris 2012}, program of Mairie de Paris. He expresses his gratitude to their hospitality and support.

\end{document}